\documentclass[11pt,onecolumn]{article}

\pdfoutput=1

\usepackage{times}
\usepackage{graphicx}
\usepackage{subcaption}
\usepackage{amsmath}
\def\upto{\mathbin{\mathstrut{\ldotp\ldotp}}}
\usepackage[text={7in,9in},centering,letterpaper]{geometry}

\begin{document}

\title{Performance analysis of IEEE 802.11ax heterogeneous network in the
presence of hidden terminals}

\author{M. Zulfiker Ali,
       Jelena Mi\v{s}i\'{c},
  and  Vojislav B.\ Mi\v{s}i\'{c}
\thanks{M. Z. Ali, J. Mi\v{s}i\'{c}, and V. B. Mi\v{s}i\'{c} are
with Ryerson University, Toronto, ON, Canada M5B 2K3,
e-mail: \{mzulfiker.ali, jmisic, vmisic\}@ryerson.ca.}%
\thanks{This report is an expanded and revised version of the paper
	`Impact of hidden nodes on uplink transmission in IEEE
	802.11ax heterogeneous network' \cite{mzali18:iwcmc}
	which has been presented at the IWCMC 2018 conference
	in Limassol, Cyprus, in June 2018.}}

\maketitle
\thispagestyle{empty}

\begin{abstract}
Performance improvement has been among the foci of all previous
amendments of IEEE 802.11 protocol. In addition, the draft high
efficiency (HE) amendment IEEE 802.11ax, proposed by TGax,   aims at
increasing network performance.  One of the main obstacles to improving
 spectral and power efficiency is the presence of hidden terminals which degrade throughput, in particular in  uplink transmission. IEEE 802.11ax does provide mechanisms such as 
trigger based uplink transmission that mitigate this degradation to some extent, but are incapable of eliminating it, esp. at high arrival rates.  To combat the hidden terminal problem, we propose to 
increase the carrier sensing threshold   (CSTH) of STAs during
association with an HE access point. Our results confirm that the proposed mechanism can lead to significant reduction of collision probability in
uplink transmission.
\end{abstract}

\maketitle

\hrule

\section{Introduction}\label{intoduction}

The ubiquitous IEEE 802.11 standard has undergone several revisions or amendments that aim to
improve its performance, in particular network throughput, by
augmenting the capabilities of its physical (PHY) and medium access
control (MAC) layers. In all cases, however, the presence of hidden nodes/terminals has
been among the major factors leading to degraded network
throughput and, consequently, spectrum inefficiency \cite{bado:2016}.  Although the
Ready-to-send (RTS)/Clear-to-send (CTS) handshake mostly succeeds in avoiding hidden terminal problem in downlink
transmission, it cannot do the same for the uplink
direction where hidden node transmissions lead to collision at the Access Point (AP). Each collision leads to retransmission and, eventually, to dropping the packet which can severely limit
the throughput of the network. 

One of the recent attempts to enhance spectral and power efficiency is the IEEE
802.11ax protocol which is specifically targeting dense WiFi deployment environments \cite{khorov:2015}
\cite{ajami:2017}. The development of this version of the standard was guided by the fact that well known techniques such as
exponential backoff, inter-frame spacing, and RTS/CTS  tend to reduce spectrum efficiency of Enhanced Distributed
Channel Access (EDCA). At the same time, the multi-user
multiple-input multiple-output (MU-MIMO) technology, which has been successful for downlink transmissions \cite{zulfiker:2017}, should be extended
to the uplink direction as well.  

Furthermore, the 802.11ax revision puts
additional focus on the coexistence of new High Efficiency (HE) devices with legacy, non-HE ones.  This is due to the long adoption time for HE protocols during which both HE and non-HE devices will coexist, and non-HE devices may experience significant performance degradation.
To ensure access fairness and backward
compatibility, the new standard will rely on EDCA
channel access in conjunction with Orthogonal Frequency Division
Multiple Access (OFDMA). As the result, both the AP and
individual stations (STAs) will compete for channel using EDCA channel access.  When the AP gains access, it will work as a central controller which will solicit CSI reports and buffer state information from peripheral STAs.  IN addition, the AP will
allocate resources for MU transmission in both uplink and downlink directions,
and initiate OFDMA random access procedure for STAs. When a non-HE
STA wins the contention, a single user transmission will take place.
This limitation has been removed in a recently proposed multi-user access protocol for uplink 
\cite{zulfikertwc:2018} which allows multiple legacy STAs to transmit
in uplink direction using MU-MIMO technique.  

To eliminate hidden
node problem in uplink MU-MIMO transmission, IEEE 802.11ax draft
protocol introduces the so-called trigger-based uplink transmission \cite{sou:2017}.
Yet even that mechanism does not succeed in eliminating hidden node interference for legacy devices, and the impact of hidden nodes in UL-MIMO transmission
remains a major concern.  This problem has been discussed, and a possible solution has been proposed, in a a recent conference paper
\cite{mzali18:iwcmc}.  The current report is a revised and extended
version of that conference paper, in which we evaluate major performance metrics of an IEEE 802.11ax
heterogeneous network in presence of hidden terminals; propose a simple solution to reduce or even eliminate the impact of hidden terminals by increasing the Carrier Sensing Threshold (CSTH) from
-82dBm to -73dBm during association of STAs with AP; and propose and discuss some modifications in the draft IEEE 802.11ax
protocol.

The rest of the paper is organized as follows: In Section
\ref{uplink}, we discuss the uplink access technique in IEEE
802.11ax amendment. Section~\ref{impact} discusses and models the impact of hidden
terminals as well as possible ways in which the impact of hidden nodes can be eliminated and the associated modifications to the draft standard specification. The simulation results are discussed in
Section~\ref{result}, followed by conclusion in Section~\ref{conclusion}.

\begin{table*}[!t]
\caption{List of abbreviations used in the
paper.\label{abbreviations}}
\begin{center}
\begin{tabular}{llll} \hline
Abbreviation & Description  & Abbreviation & Description
\\ \hline 
AC     & Access Category & AIFS & Arbitration Inter-frame Spacing\\                    
A-MSDU & Aggregate MSDU& BO&Backoff \\
CSI    & Channel State Information & CTS & Clear to Send\\ 
CW     & Contention  Window    & DCF & Distributed Coordination Function\\
DIFS   & DCF Inter-frame Spacing& EDCA         & Enhanced Distributed Channel Access \\ 
EDCAF  & EDCA Function       & G-ACK & Group ACK\\
G-CTS  & Group CTS& HPG & High Priority  traffic Group\\
LPG    & Low priority  traffic Group& MIMO         & Multiple Input Multiple Output      \\ 
MPDU   & MAC Protocol Data Unit     & MPR & Multi Packet Reception\\
MSDU   & MAC Service Data Unit & MU           & Multi User                          \\
NDP    & Null Data Packet& NDPA & NDP Announcement\\
PPDU   & PHY Protocol Data Unit& RTS & Ready to Send\\
SIFS   & Short Inter-frame Spacing  & VHT & Very High Throughput\\
TXOP   & Transmission opportunity  & AIFSN  &AIFS number Throughput\\ 
LST    & Laplace-Stieltjes transform  & NDP  &Null Data Packet\\ 
PGF    & Probability generating function  &  MU-MIMO & Multi-user MIMO\\
DL-MIMO & Downlink MIMO  &  OFDMA & Orthogonal frequency division multiple access\\
A-MPDU & Aggregate MPDU  & STA & Station \\
AP     & Access point  &  &  \\ \hline
\end{tabular}
\end{center}
\end{table*}

\section{Uplink transmission in IEEE 802.11ax protocol}\label{uplink}

In an IEEE 802.11ax-compliant network, multi-user uplink transmission is performed under control of the 
AP. A new control frame format, known as the trigger frame, contains the 
information that identifies the STAs that will transmit uplink multi-user
PHY layer packets, hereafter referred to as PHY layer protocol data units
 (PPDUs).  Trigger-based uplink transmission requires the AP to have the
buffer state information for all STAs that intend to transmit. This information can be supplied by the STAs themselves (although the specification does not define how exactly is this to be done), or the AP can explicitly request it by
transmitting a trigger frame.  Nonetheless, we assume that the STA must access the medium through EDCA procedure, even if only to transmit buffer state
information to the AP.  

\begin{figure*}[t!]
    \centering
    \begin{minipage}[t]{0.475\textwidth}
        \centering
        \includegraphics[width=0.95\textwidth]{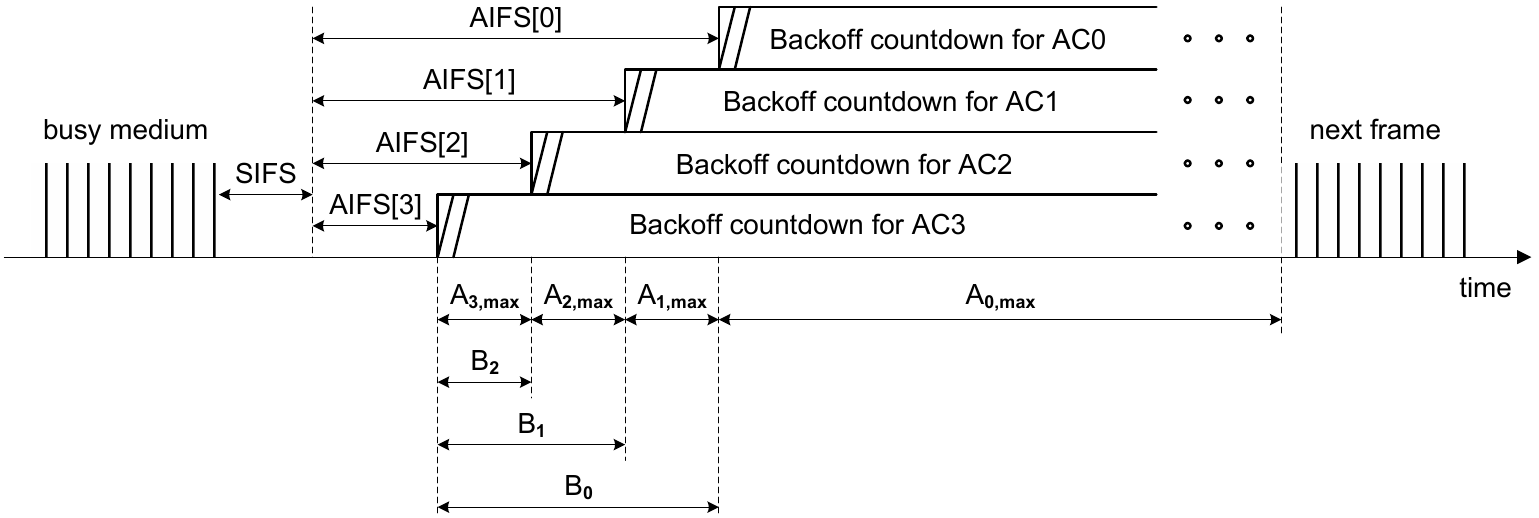}
        \caption{EDCA channel prioritized access (adopted from
\cite{misic:2012}).\label{model}}
    \end{minipage}%
    ~ 
    \begin{minipage}[t]{0.5\textwidth}
        \centering
        \includegraphics[width=0.95\textwidth]{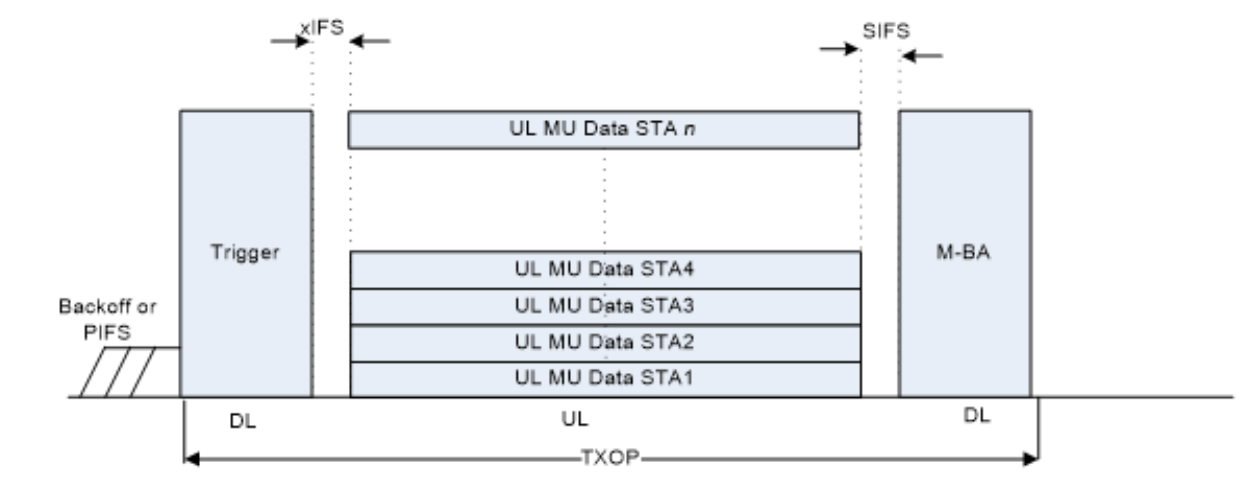}
        \caption{Trigger-based UL MU transmission (adopted from \cite{standard:11ax}).\label{trigger}}
    \end{minipage}
\end{figure*}

An UL MU transmission is initiated as an immediate
response to a DL trigger frame sent by the AP, as shown in Fig.~\ref{trigger}. The trigger frame allocates resources for the intending
STAs. The frame exchange initiated by the trigger frame is considered successful if the AP receives UL MU
data correctly from at least one STA. AP then acknowledges the
successful receipt of all MPDUs by sending the so-called Multi-block acknowledgement
(M-BA). 


Before responding to a trigger frame, the STA performs physical (ED) as
well as virtual (NAV) carrier sensing if trigger frame indicates to do
so before UL MU transmission \cite{kiseon:2016}.

The multi-user uplink transmission protocol for legacy devices is
discussed in detail in \cite{zulfikertwc:2018}.

\section{Impact of hidden nodes} \label{impact}

\begin{figure*}[t!]
    \centering
    \begin{minipage}[b]{0.4\textwidth}
        \centering
        \includegraphics[width=0.95\textwidth]{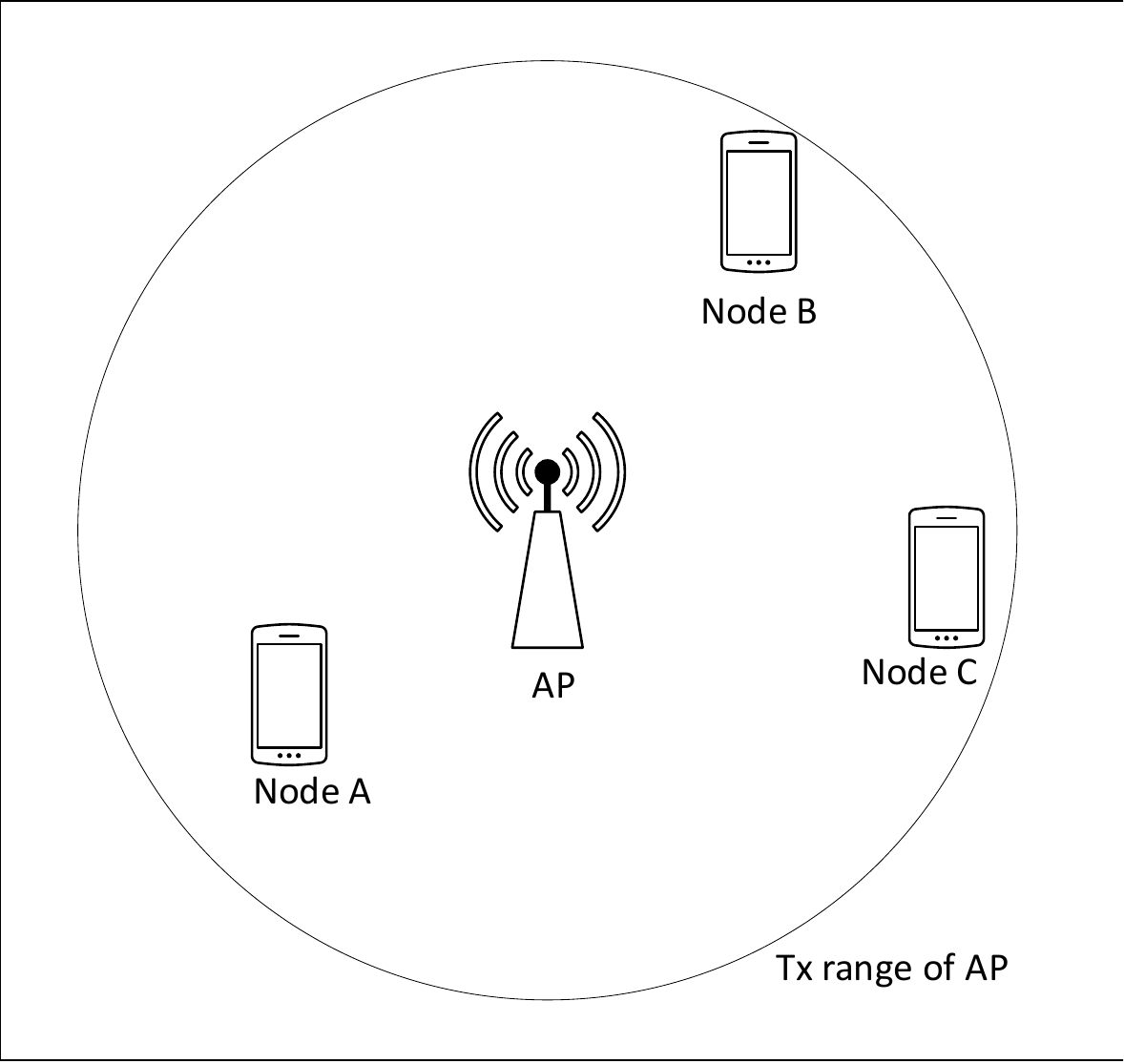}
        \caption{Basic Service Subset.\label{bss}}
    \end{minipage}%
    ~ 
    \begin{minipage}[b]{0.55\textwidth}
        \centering
        \includegraphics[width=0.95\textwidth]{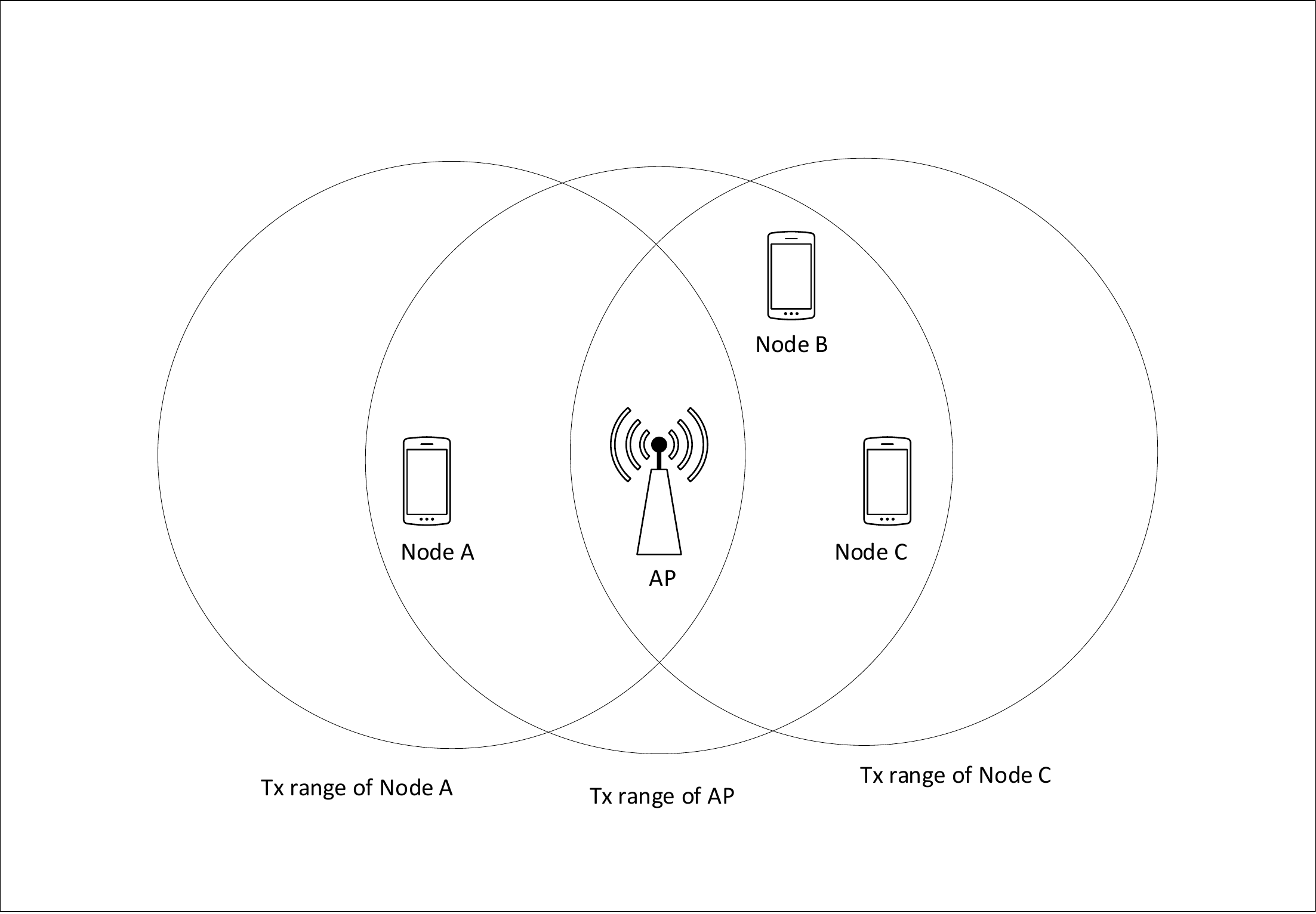}
        \caption{Hidden Node problem.\label{hiddennode}}
    \end{minipage}
\end{figure*}

To discuss the impact of hidden nodes, let us consider a heterogeneous network consisting of only
one HE AP and a number of HE and legacy (i.e., non-HE) STAs in an isolated basic
service subsystem (BSS) as shown in Fig.~\ref{bss}. Assuming that other BSS(s) in the vicinity operate on different primary channel(s), 
there will be no interference from the nearby BSS. Since all the nodes are associated with the AP, we can
safely assume that all the nodes are within the transmission range of
the AP. Therefore, there are no hidden
terminals from the AP point of view and no associated hidden terminal problem. However, in the AP-initiated uplink multi-user
communication, there may be concurrent transmissions from other STAs, which means that the trigger frame itself is
vulnerable to collision. In other words, the only transmission initiated
by the AP that may suffer a collision is the trigger frame. 
In addition, as soon as the
STAs receive trigger frame or MU-RTS, they will update their NAVs
accordingly.  Therefore, data, CTS or
ACK/G-ACK messages from HE devices transmitted in the uplink direction will not suffer any collision.  

However, due
to the coexistence of legacy devices, uplink SU or MU transmission from
legacy devices cannot avoid the hidden terminal problem.  As shown in Fig.~\ref{hiddennode}, when a node (A) initiates an RTS in the uplink, assuming that the transmission range of the APP and all other nodes is the same, nodes B and
 C effectively act as hidden terminals for node A. Therefore, any ongoing RTS
transmission between node A and AP is not noticed by node B or node C.
If another node, say, node C, initiates a transmission during an ongoing RTS
transmission between node A and AP, or within the SIFS period before the CTS
transmission by AP, the AP will see those packets as collisions.
As the result, both node A and node C will not receive their CTS from AP
which will cause both nodes A and C to increase their contention windows and initiate 
retransmissions.

\subsection{Modeling hidden nodes}\label{modeling}

To model the impact of hidden nodes, we will assume that STA traffic can be categorized in four priority categories
($k=0 \upto\ 3$).  Furthermore, we assume that each traffic category will 
contend for medium access using its own EDCA function. Packet arrival for
traffic class $k$ is assumed to follow a Poisson process with arrival rate of
$\lambda_k$.  Durations of MPDU, RTS, CTS, BA and Trigger frames, denoted by
$l_d, rts, cts, ba,$ and $trig$, respectively, are expressed as
integer multiples of slots (we assume a time slot of $\omega =
9\mu$s). Durations of $AIFS_k, k= 0 \upto 3$ and SIFS periods are
denoted by $aifs_k$ and $sifs$, respectively (in slots). Channel state is
modeled via bit error rate $ber$ so that the probability that a RTS
or CTS frame will not be corrupted by channel noise is $\delta =
(1-ber)^{rts_b+cts_b}$ .  Probability that data or BA frames  will not
experience any corruption due to channel noise is denoted as $ \sigma
=(1-ber)^{l_{db}+ba_b}$. In the two last equations, $rts_b, cts_b, ba_b$ and $l_{db}$ denote the bit count in RTS, CTS, BA and MPDU frames, respectively.

The behaviour of different ACs during backoff under EDCA is shown in
Fig.~\ref{model}.  Before $AC_k$ can begin backoff countdown, the
medium must be idle for a period of $AIFS_k$ without interruption.
$AIFS_k$ can be expressed as $AIFS_k = SIFS + AIFSN_k\omega $ where
$AIFSN_k$ is arbitration inter frame spacing number for traffic category
$k$.  Since priority increases with $k$, no transmission is possible
during the period $AIFS_3$. Initial values of the freezing
counters are set to be $B_k = AIFSN_k - AIFSN_3, k = 0 \upto 3$.

Duration of time periods in which traffic class $k$ and higher can access the
medium is denoted as $A_k$ and their maximum values are
\begin{equation}
A_{k,max} = \left\{
\begin{array}{lr}
\!\!(AIFSN_{k-1} - AIFSN_k) \omega, & k = 1 \upto 3 \\
\!\!W_{0,max},                      & k = 0
\end{array} \right.
\end{equation}
where $W_{0,max}$ is the maximum number of backoff states for $AC_0$.

The initial value of the backoff counter is uniformly distributed over the interval $0
\upto CW_{k,i}$, where $CW_{k,i}$ is the contention window for $AC_k$
during backoff phase $i$.  The maximum number of backoff states in
backoff phase $i$ is $W_{k,i}$ = $CW_{k,i}+ 1$.  For backoff phase 0,
the number of backoff states $W_{k,0}$ is set to $W_{k,0} =
CW_{k,min}+1$. We can express $W_{k,i}$ in terms of $W_{k,0}$ as
\begin{equation}
W_{k,i} = \left\{
\begin{array}{lr}
2^i W_{k,0}                 & 0 \leq i, \leq m_k \\
2^{m_k} W_{k,0} = W_{k,max} & m_k < i \leq R
\end{array} \right.
\end{equation}

If $N_{k,t}$ denotes the number of nodes within the transmission range of a
STA, we can relate the probability $f_k$ that a time slot will be idle during
the interval $A_{k,max}$ and the channel access
probability $\tau_k$ as
\begin{align}
f_k =  \prod_{l=k}^{3} (1-\tau_l)^{N_{k,t}}
\end{align}

Probability of successful transmission for a node, without considering
the impact of hidden nodes, can be written as
\begin{align}\label{gk} 
\gamma_0 &= \frac{f_0}{1-\tau_0} \nonumber\\
\gamma_1 &=  (1\!-\!{f_1}^{A_{1,max}})\frac{f_1}{1\!-\!\tau_1}
+ {f_1}^{A_{1,max}}\frac{f_0}{1-\tau_1}\nonumber\\
\gamma_2 &=  (1\!-\!{f_2}^{A_{2,max}})\frac{f_2}{1\!-\!\tau_2}  \nonumber\\
&+ {f_2}^{A_{2,max}} 
\left[ (1\!-\!{f_1}^{A_{1,max}}) 
\frac{f_1}{1-\tau_2} + {f_1}^{A_{1,max}}
\frac{f_0}{1-\tau_2}\right]\nonumber\\
\gamma_3  &=  (1\!-\!{f_3}^{A_{3,max}})\frac{f_3}{1\!-\!\tau_3}
+ {f_3}^{A_{3,max}}\left\{ (1\!-\!{f_2}^{A_{2,max}})
\frac{f_2}{1\!-\!\tau_3} \right.\nonumber\\
&\left.+ {f_2}^{A_{2,max}}\left[ (1\!-\!{f_1}^{A_{1,max}})
\frac{f_1}{1\!-\!\tau_3}
+ {f_1}^{A_{1,max}}\frac{f_0}{1\!-\!\tau_3}\right]\right\}
\end{align}

The vulnerable period, i.e., the period during which a collision can take place
at the AP, for a single user uplink transmission is $T_{v-su}=(rts + sifs
+ cts)$. By the same token, the vulnerable period for an UL MU transmission is
$T_{v-mu}=trig$. 

Let $N_{k,h}$ be the number of nodes of traffic category $k$ hidden from 
a transmitting node. The probability that a hidden node
will not transmit in any time slot during vulnerable period can be
written as
\begin{equation}
f_{h} =  \sum_{k=0}^{3}\frac{(1- \tau_k)^{N_{k,h}}}{\tau_kN_{k,h}}
\end{equation}
where $\tau_k$ is the transmission probability for traffic category
$k$.

Then, the probability of no collision from the
hidden nodes during the entire vulnerable period is
\begin{equation}
f_{ncoll} =  f_h^{(1-f_{mu})T_{v-su}+f_{mu}T_{v-mu}}
\end{equation}
where $f_{mu}$ is the trigger transmission probability of AP to
initiate multi-user uplink transmission. This probability is equal to the
downlink MU TXOP sharing probability given by
\begin{align} \label{txopshare}
f_{mu} =1-\frac{ Th_k}{\sum_{m=0}^{3}Th_m}
\end{align}
as discussed in \cite{zulfikertwc:2018}.

Finally, probability of successful transmission in the presence of hidden
nodes can be expressed as
$\gamma_{k,h}=\gamma_kf_{ncoll}$, where $\gamma_k$ is the probability of successful
transmission but without any hidden node.

\subsection{Eliminating the hidden node problem}\label{approach}

Due to rapid growth of diverse and dense deployment environments, IEEE 802.11ax networks can be expected to contain an ever increasing number (and, consequently) density of both APs and STAs. These environments are commonly characterized by
hidden terminal problems and the resulting increase in interference from the nearby WLANs, increase in collision rate, and decrease of channel utilization.  A prpmising technique seems to be to increase the carrier sensing
threshold (CSTH) which will allow
multiple STAs and/or APs to transmit simultaneously,
which will counter the performance degradation caused by collisions and improve spatial reuse.  At the same time, increasing CSTH in a coexistence scenario may lead
to asymmetric hidden node problem in which legacy STAs will be severely
deprived of the chances of channel access due to the HE STAs virtually taking over the shared channel \cite{mvulla:2015}.

\begin{table*}[!t]
\caption{Parameters used for simulation.\label{parameter}}
\begin{center}
\begin{tabular}{ll} \hline
Parameters & Numerical values  \\
\hline
Bit error rate, BER & $2X10^{-6}$ bits/s \\
Duration of Time slot, $\sigma$ & 9 $\mu s$\\
Minimum PHY header & 40 $\mu s$ \\
Maximum PHY header & 52 $\mu s$\\
DCF Inter-frame space duration, DIFS & $34  \mu s$  \\
MPDU length  & 11454 \\
Short Inter-frame space duration, SIFS & $16  \mu s$ \\
MAC header length  & 36 bytes\\ 
Request to send, RTS & 20 bytes \\
Clear to send, G-CTS & 14 bytes\\
Acknowledgement,  G-ACK & 32 bytes \\
Maximum retry limit , R & 7\\ 
Max. number of antennas  in AP, $M_{ant}$  & 4  \\
Number of antennas  in STA  & 1\\ 
Bandwidth  & 80 MHz \\
OFDM symbol duration & 4 $\mu s$\\
Number of bits per OFDM symbol duration  &  1560 \\
Modulation and Coding scheme, MCS  & 9  \\ 
Maximum backoff stages & $ [5,5,1,1] $ \\
Arbitration inter frame space, AIFS  & $[ 7,5,3,2 ]$\\ 
Minimum contention window size $CW_{min} $ &  $[ 32,32,16,8 ]$ \\
TXOP  duration limit & $ [0,0,1504,1504] \mu s $ \\
\hline 
\end{tabular}
\end{center}
\end{table*}

We argue that the hidden terminal problem can be solved by increasing
the carrier sensing threshold (CSTH) of all STAs during association
with a HE AP.  This will effectively reduce the transmission range of the AP and restrict the number of STAs that can associate with the AP to those physically closer to the AP.
However, during usual transmission process, the carrier sensing
threshold will remain at the value recommended by the standard for all STAs.
In this way we can ensure that all STAs within the BSS are also within
the transmission range of each other. 

This approach allows us to solve the hidden node problem in the uplink transmission and, furthermore, to reduce the transmit power of the AP which will make the AP
more energy efficient. However, it may also lead to inter-BSS
interference issue.  Namely, if a STA has a larger transmission range, it will
generate interference to the STAs of the nearby BSS and even attempt to set the
NAV of the STAs belonging to other BSS. Fortunately, this interference can easily
be eliminated if the adjacent BSSs are forced to operate on different
primary channels and if we restrict the transmission from a foreign BSS in
setting the NAV of a STA. 

In light of this discussion, we propose the following
modifications in the draft specification:

\begin{itemize}

\item The CSTH of all STAs during association with HE AP will be
increased while CSTH of all STAs during normal transmission process
will remain same as legacy STAs.

\item Two adjacent BSSs will operate in two different primary
channels.

\item The NAV of a STA will be set only by the transmission of another
STA within the same BSS. If a STA receives transmission from another
STA which does not belong to same BSS as the receiving STA, NAV will
not be updated.

\end{itemize}

\section{Simulation Result and Discussion}\label{result}

To
simulate the MAC layer of IEEE 802.11ax protocol, we have developed an event-driven simulator in Matlab.

\begin{figure*}[t!]
    \centering
    \begin{minipage}[b]{0.475\textwidth}
        \centering
        \includegraphics[width=0.95\textwidth]{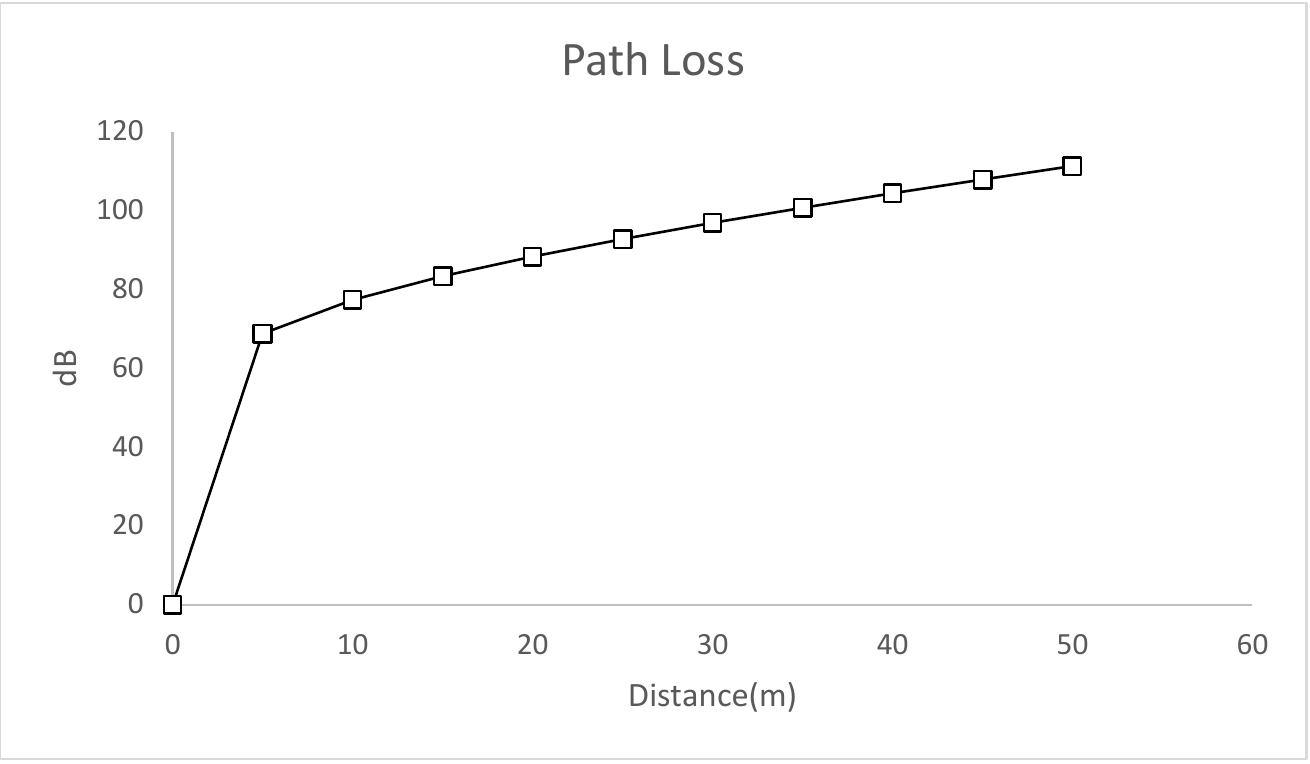}
        \caption{Path Loss (proposed 11ax standard).\label{pathloss}}
    \end{minipage}%
    ~ 
    \begin{minipage}[b]{0.475\textwidth}
        \centering
        \includegraphics[width=0.95\textwidth]{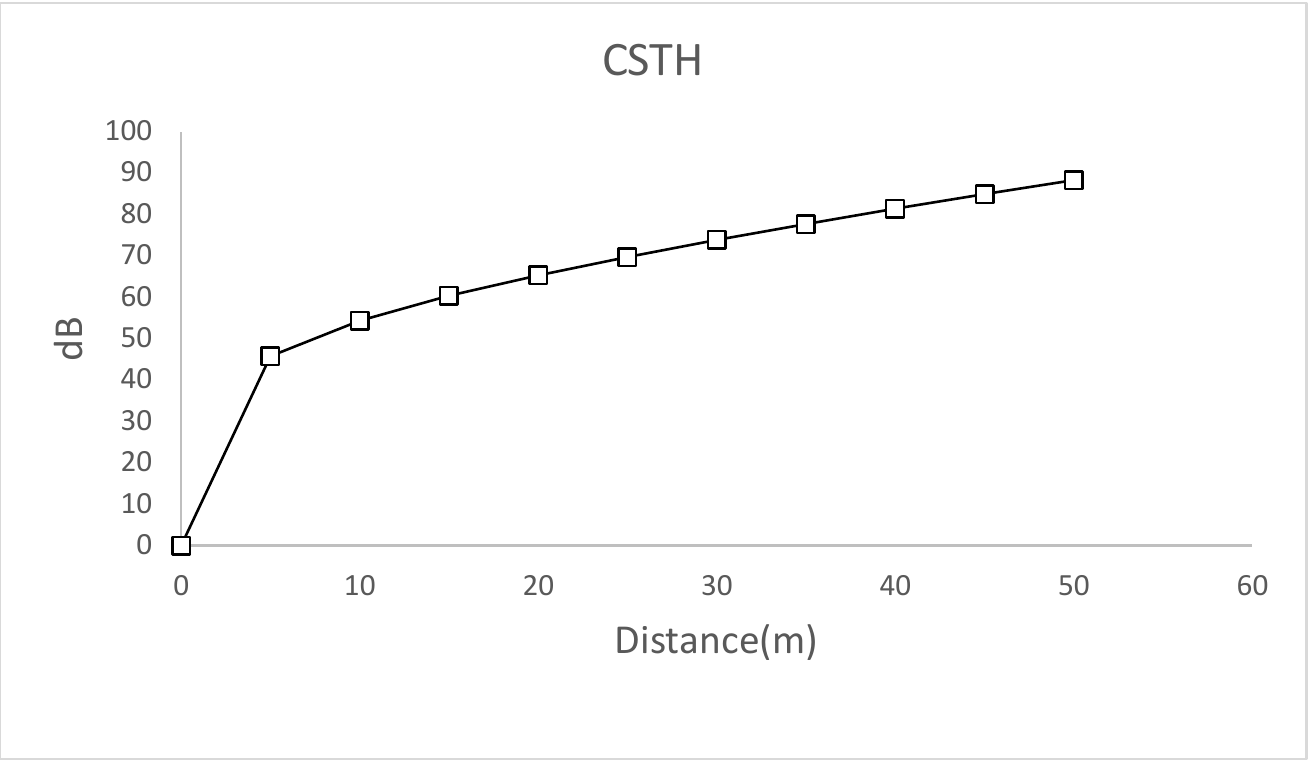}
        \caption{Transmission distance for different CSTH.\label{csth}}
    \end{minipage}
\end{figure*}

We assume an indoor
environment with a single AP and a number of STAs placed randomly on the
same floor.  We have used the 3GPP indoor femto pathloss (PL) model,
as recommended for IEEE 802.11ax standard, as shown in Fig. \ref{pathloss} where pathloss
exponent is 2, a linear attenuation for walls is 0.5 dB/m, and 4 dB is assumed to model the shadowing effect in indoor environment.  

We assume
transmit power level of 23 dBm (200 mW) and initial carrier sensing
threshold of -82 dBm.  The concentration of the nodes is controlled so that the AP has always 24 nodes associated with it.  For each
simulation setup, when we increase the carrier sensing threshold of
STAs during association process, the transmission range of AP is actually
reduced. Therefore, we increase the concentration of STA to
ensure that we have same number of STAs associated with AP. 

We further
assume that the CSTH of all STAs during usual transmission process
remains -82 dBm.  

The simulation is run with uniformly
varying load intensity for each node and all traffic categories. Each simulation is
run for 1 second and the performance metrics are averaged over 10
runs. 

We assume that our VHT PPDU has one MPDU which contains an
A-MSDU originating from the LLC layer. The size of the A-MSDU is limited by the
maximum size of MPDU which is 11454 octets. Throughout the
evaluation we treat an MPDU as a packet so that the arrival of a  packet means
the arrival of an MPDU to the queue.  

Our network has a bandwidth of 80
MHz having 234 usable subcarriers. A 256 QAM modulation with 5/6
coding scheme allows 1560 bits to be transmitted per OFDM symbol
duration of 4 $\mu s$ with long guard interval.

Other pertinent parameters for the model are shown in
Table \ref{parameter}.

In this setup, we
evaluate the performance of the network in three different scenarios.
In the first case, we look at the performance of the network under varying packet arrival rate in the presence of hidden terminals. In the
second case, we vary the
packet arrival rates without hidden terminals.  Lastly, we gradually
increase the carrier sensing threshold during the association process for
a specific packet arrival rate (4800 packets/sec) and evaluate the
performance of the network.

\begin{figure*}[t!]
    \centering
    \begin{minipage}[b]{0.33\textwidth}
        \centering
        \includegraphics[width=0.95\textwidth]{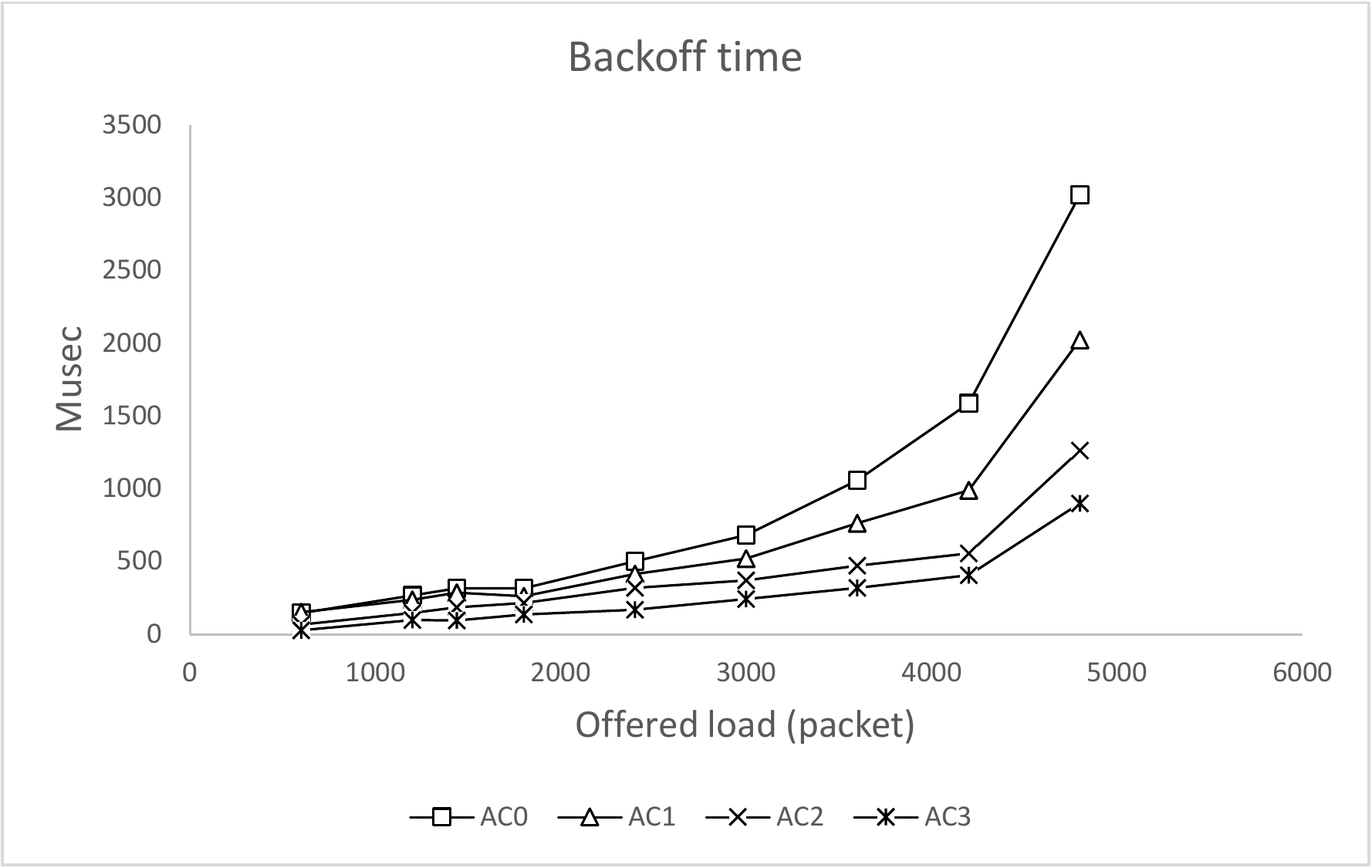}
        \caption{Backoff time, hidden node present.\label{backoffh}}
    \end{minipage}%
    ~ 
    \begin{minipage}[b]{0.33\textwidth}
        \centering
        \includegraphics[width=0.95\textwidth]{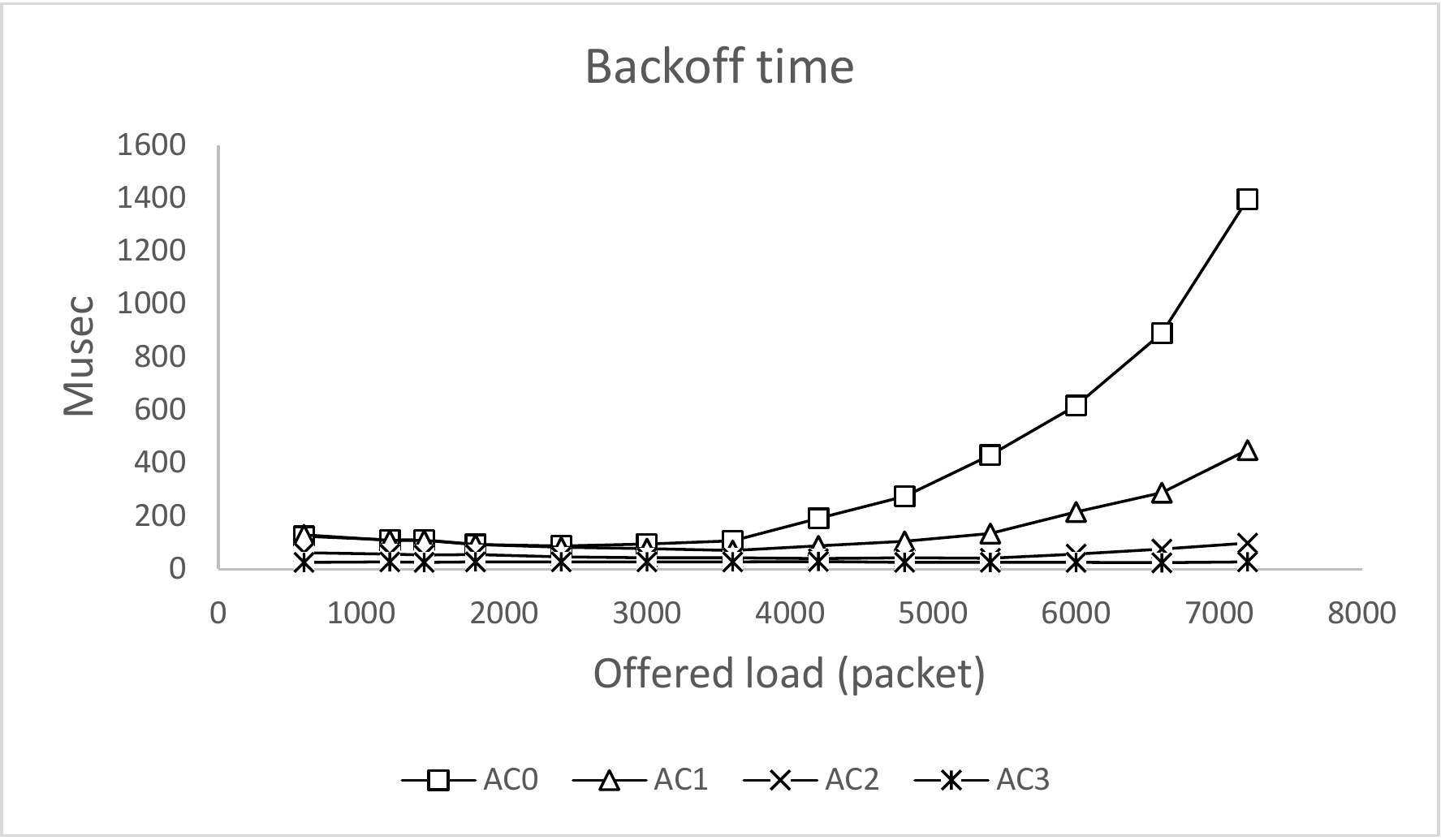}
        \caption{Backoff time, no hidden node present.\label{backoffn}}
    \end{minipage}%
    ~
    \begin{minipage}[b]{0.33\textwidth}
	\centering
	\includegraphics[width=0.95\textwidth]{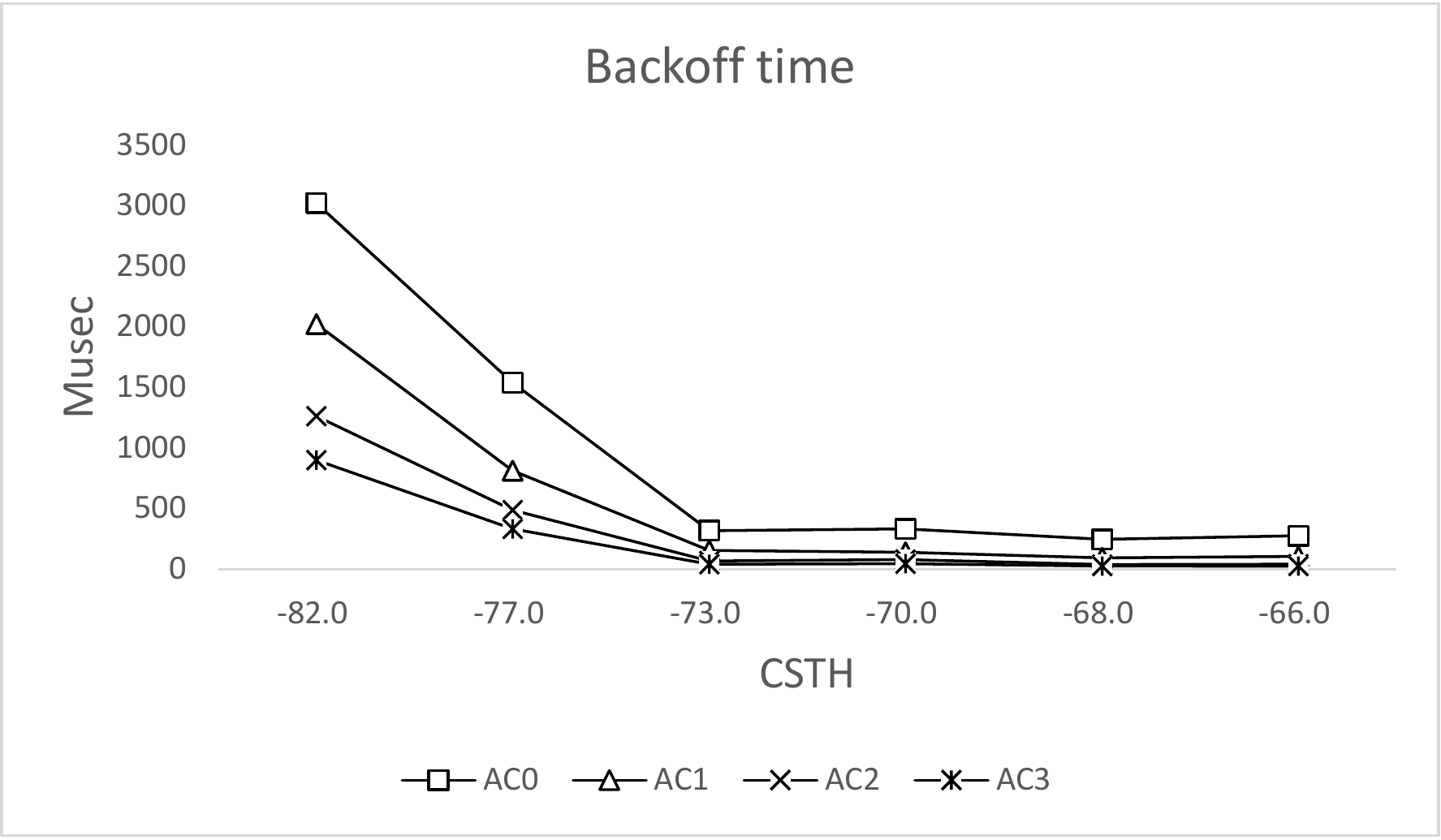}
	\caption{Backoff time with CSTH increased to 4800
	packets/sec.\label{backoffcsth}}
    \end{minipage}
\end{figure*}

Figs.~\ref{backoffh}, \ref{backoffn}, and \ref{backoffcsth} shows the backoff time for the scenarios outlined above. 
The backoff times for all traffic categories gradually
increase with the increase of packet arrival rates in presence of
hidden terminal as shown in Fig.~\ref{backoffh}. Although multiple
uplink transmissions take place, due to collision
from hidden terminals we observe a steady increase in backoff time.

However, in the absence of hidden terminals, the backoff time
gradually decreases slightly at low load condition (up to 3600
packets/sec) as shown in Fig.~\ref{backoffn}. This decrease is due to
multi-user uplink transmission. During multi-user uplink transmission,
Namely, when a primary STA gets a transmission opportunity (TXOP), secondary STAs
share the TXOP with the primary STA even if those secondary STAs are in
backoff process. Those transmission opportunities of secondary STAs
reduce the effective backoff time as more and more secondary STAs get
transmission opportunities without going through the full backoff process,
and consequently average backoff time decreases. Since only four STAs
can transmit at a time in uplink direction, with the increase of
packet arrival rate (above 3600 packets/sec), many STAs need to wait
for the medium for a longer time and consequently the backoff time
increases, as shown in Fig.~\ref{backoffn}.  

At a packet arrival rate
of 4800 packets/sec, the backoff time in the presence of hidden terminals
is almost 10 times higher than without them. Also the backoff time for lower priority traffic
increases at a faster rate than that of higher priority
traffic because, at higher packet arrival rate, more higher
priority traffic gets the TXOP sharing opportunity and lower priority
traffic needs to wait longer to access the medium. 

Finally, the variation
of backoff time with the increase in carrier sensing threshold is
shown in Fig.~\ref{backoffcsth}.  We observe that at CSTH of -73dBm,
the backoff time of all traffic categories are same as the backoff
time of STAs without hidden terminals.

\begin{figure*}[t!]
    \centering
    \begin{minipage}[b]{0.33\textwidth}
        \centering
        \includegraphics[width=0.95\textwidth]{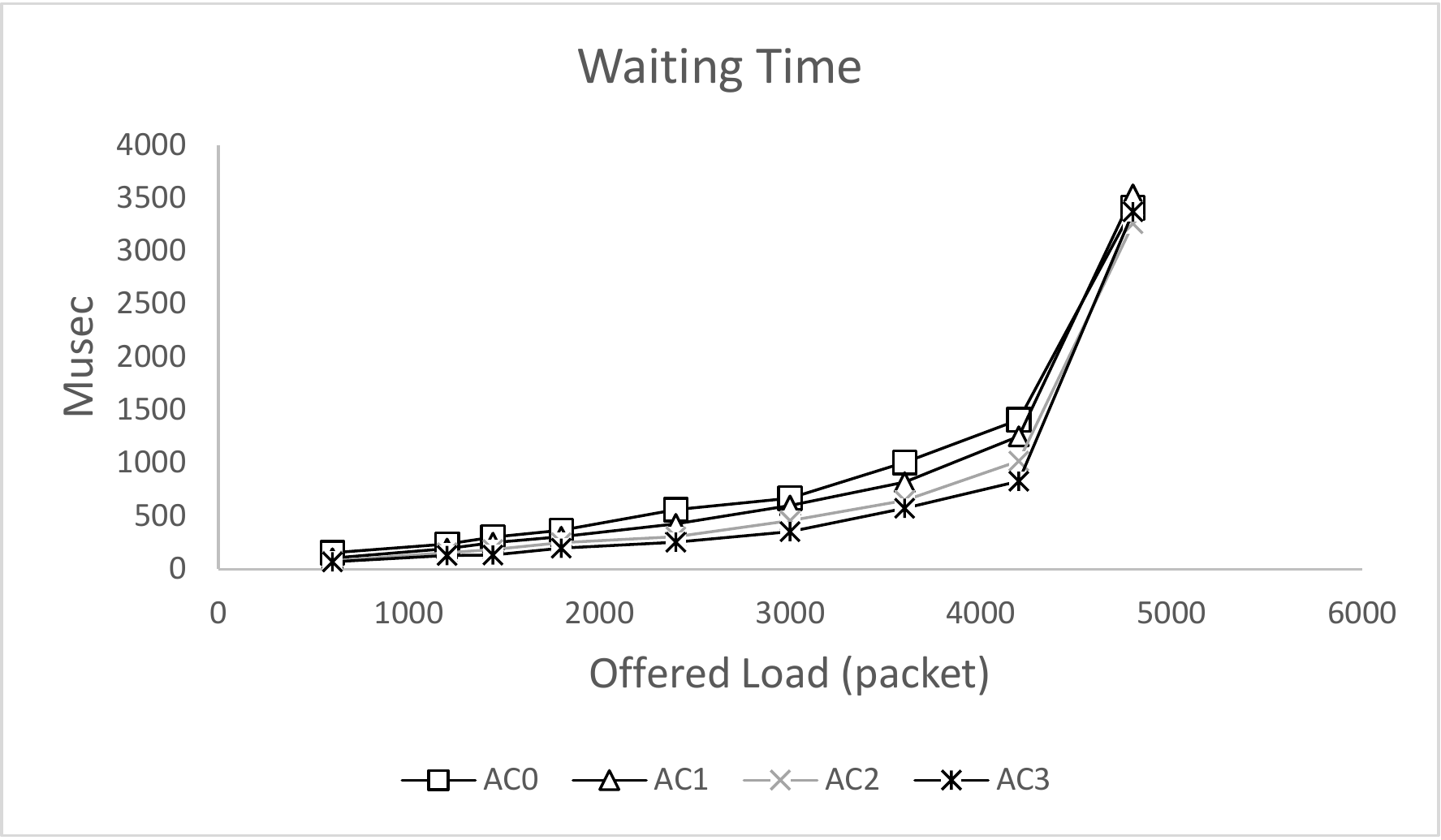}
        \caption{Waiting time in the queue, hidden node
	present.\label{waitingtimeh}}
    \end{minipage}%
    ~ 
    \begin{minipage}[b]{0.33\textwidth}
        \centering
        \includegraphics[width=0.95\textwidth]{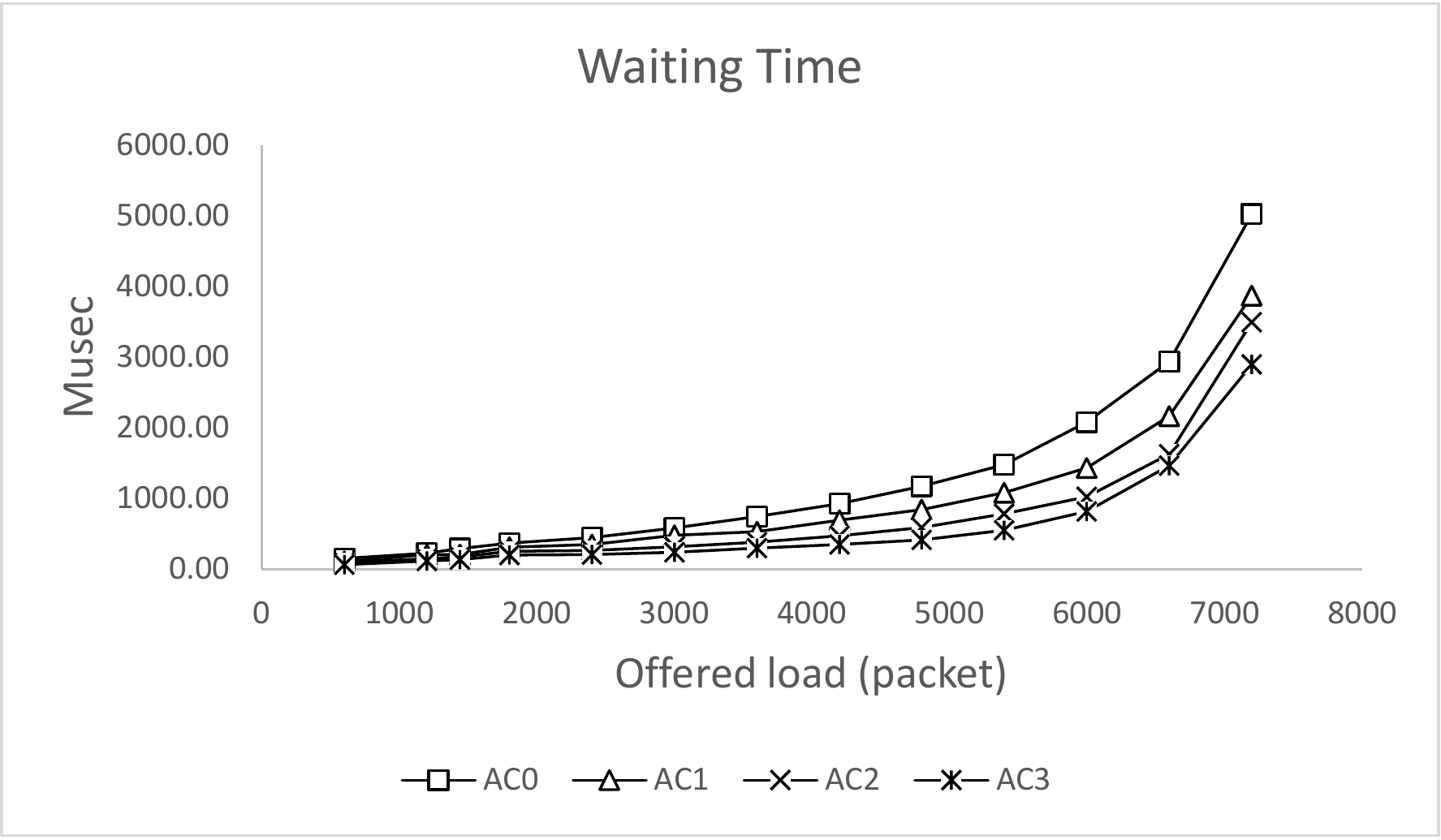}
        \caption{Waiting time in the queue, no hidden node
	present.\label{waitingtimen}}
    \end{minipage}%
    ~ 
    \begin{minipage}[b]{0.33\textwidth}
        \centering
        \includegraphics[width=0.95\textwidth]{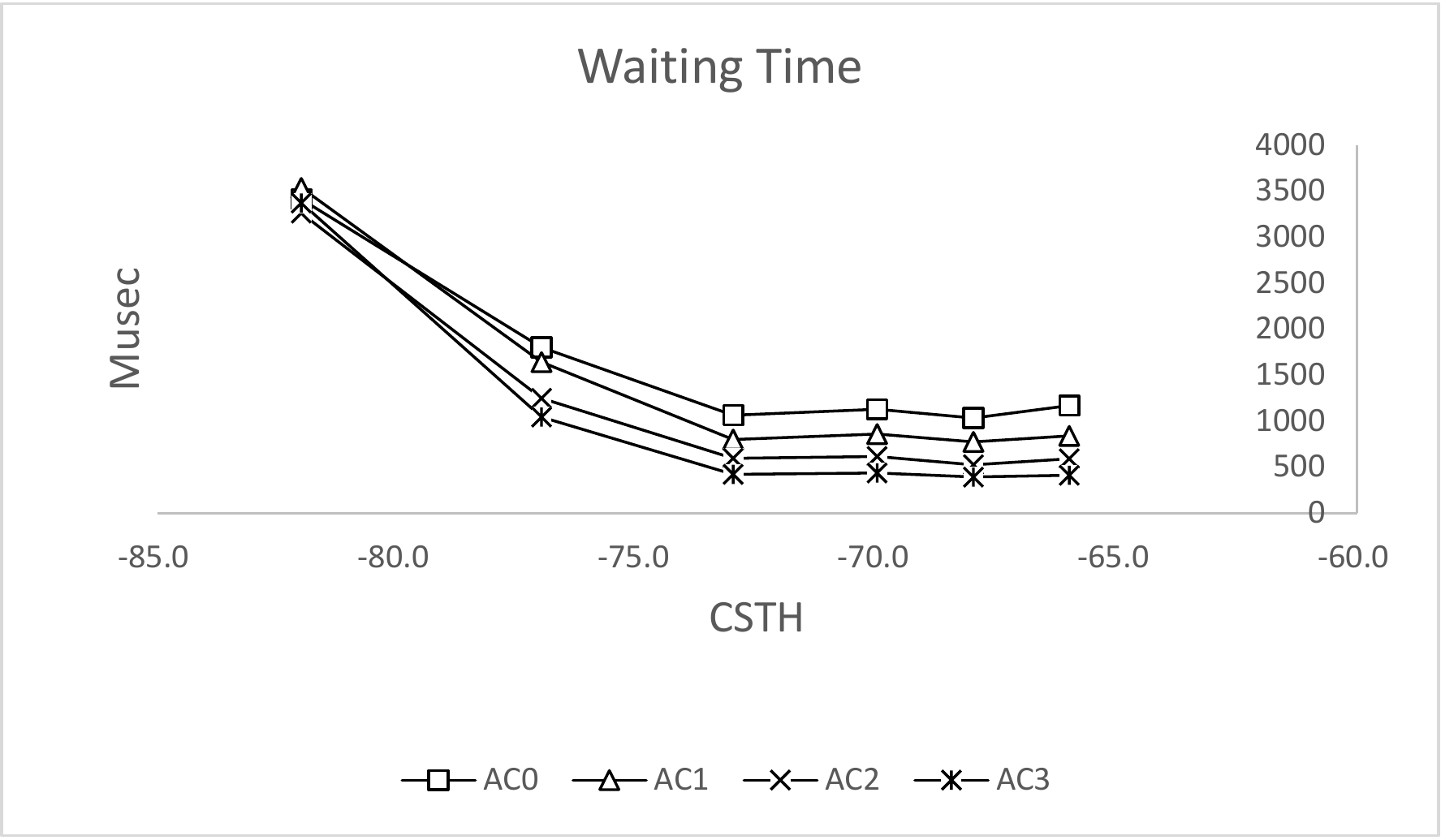}
        \caption{Waiting time in the queue with CSTH increased to 4800
	packets/sec.\label{waitingtimecsth}}
    \end{minipage}
\end{figure*}

Waiting time is the time a packet needs to wait in the queue before
starting the backoff process. Figs.~\ref{waitingtimeh}, \ref{waitingtimen}, and \ref{waitingtimecsth} show the waiting
time in three different scenarios. As shown in Figs.~\ref{waitingtimeh}
and \ref{waitingtimen}, the waiting time gradually increases with the
increase of packet arrival rate. However, the waiting time of a packet
in presence of hidden terminal is much higher than the waiting time of
a packet without hidden terminals. The decrease of waiting time with
the increase of CSTH is shown in Fig.~\ref{waitingtimecsth}.

\begin{figure*}[t!]
    \centering
    \begin{minipage}[b]{0.475\textwidth}
        \centering
        \includegraphics[width=0.95\textwidth]{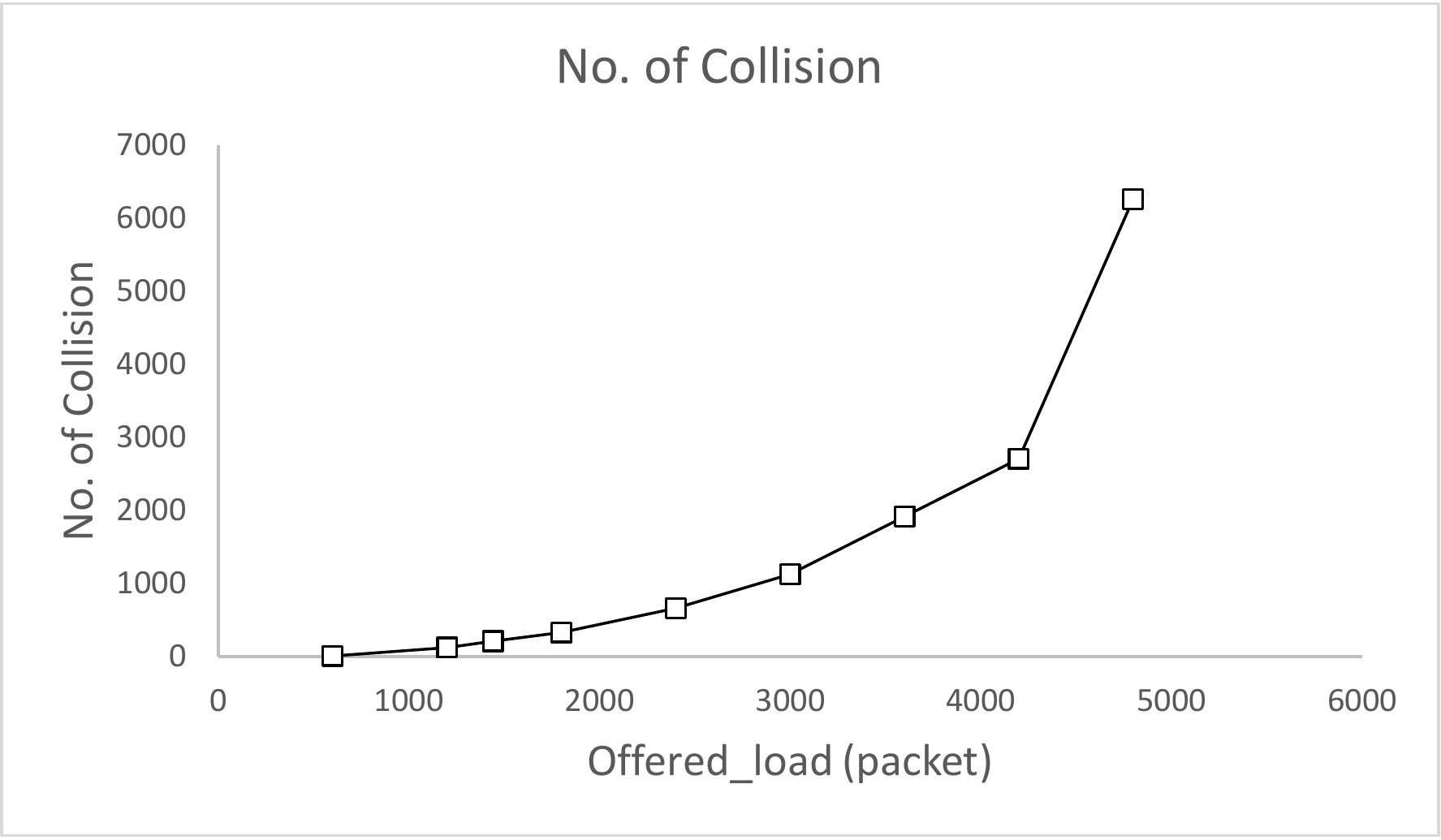}
        \caption{Number of collisions, hidden node
	present.\label{collisionh}}
    \end{minipage}%
    ~ 
    \begin{minipage}[b]{0.475\textwidth}
        \centering
        \includegraphics[width=0.95\textwidth]{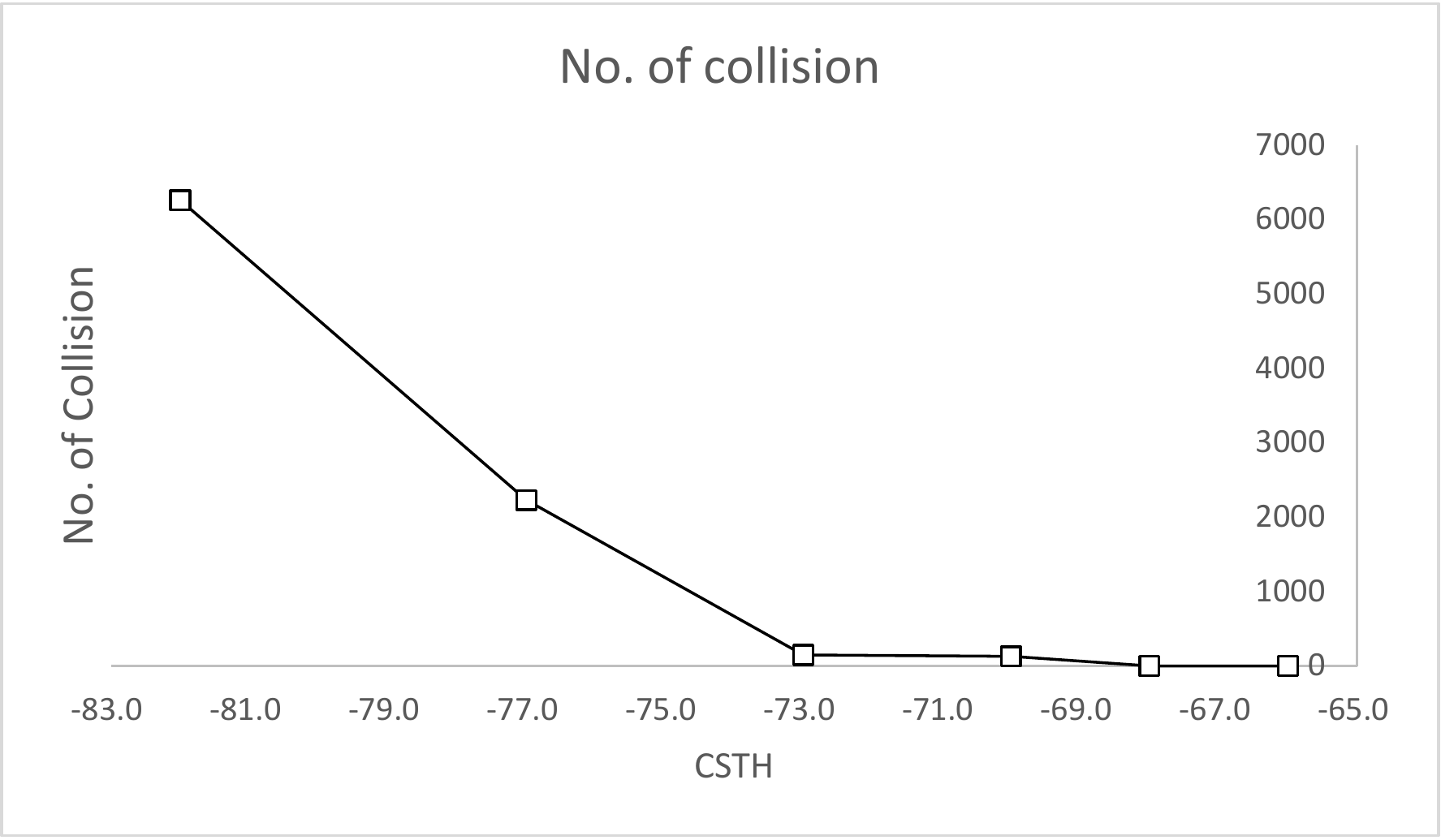}
        \caption{Number of collisions with CSTH increased to 4800
	packets/sec.\label{collisioncsth}} 
    \end{minipage}
\end{figure*}

The number of collisions due to hidden terminal(s) for an existing CSTH is shown
in Fig.~\ref{collisionh}. At low packet arrival rate, collision
probability is very low on account of longer inter packet arrival
time and SU packet transmission.  In this case, two transmissions from nodes which are hidden to each other
rarely overlap.  In addition, vulnerable time for collisions due to
hidden node is shorter for SU transmission, and inter-packet arrival
time is long enough to complete the transmission of RTS/CTS signaling.

When the packet arrival rate increases, the inter-packet arrival time
decreases and more transmissions overlap which increases the collision rate at the AP.  Collisions of RTS/CTS frames with transmissions from hidden nodes lead to retransmissions which increases the collision rate even further.
This effect may be observed even at higher packet arrival rate, where probability of simultaneous
transmission from the hidden node groups and, consequently,
the collision probability both become very high.  

Fig.~\ref{collisioncsth}
shows the number of collisions as the function of
carrier sensing threshold of STAs during the association process.  As CSTH increases, the number of collision decreases.
If CSTH is set at -73dBm, the number of collisions becomes very low; still, from Fig.~\ref{csth} we observe that the corresponding transmission
radius of AP is around 30 meters. This observation confirms our 
argument that in a densely deployed environment, the concentration of
APs need to be very high so that all nodes within the transmission
range of an AP can hear each other.  It also confirms the validity of our approach, namely that  increasing the CSTH of STAs during
association with AP will reduce or even eliminate the impact of hidden nodes.

\begin{figure*}[t!]
    \centering
    \begin{minipage}[b]{0.475\textwidth}
        \centering
        \includegraphics[width=0.95\textwidth]{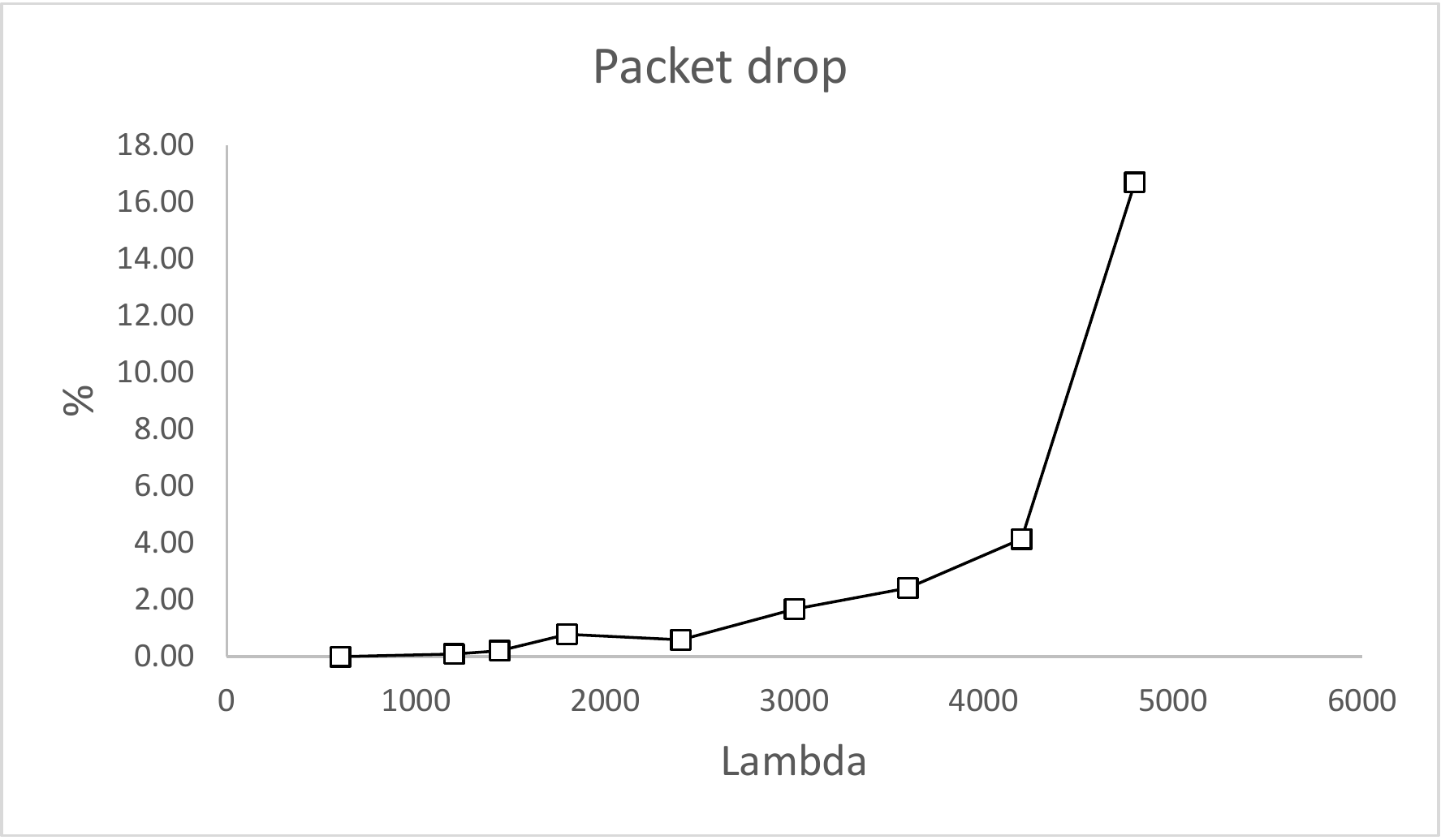}
        \caption{Packet drop, hidden node present.\label{packetdroph}}
    \end{minipage}%
    ~ 
    \begin{minipage}[b]{0.475\textwidth}
        \centering
        \includegraphics[width=0.95\textwidth]{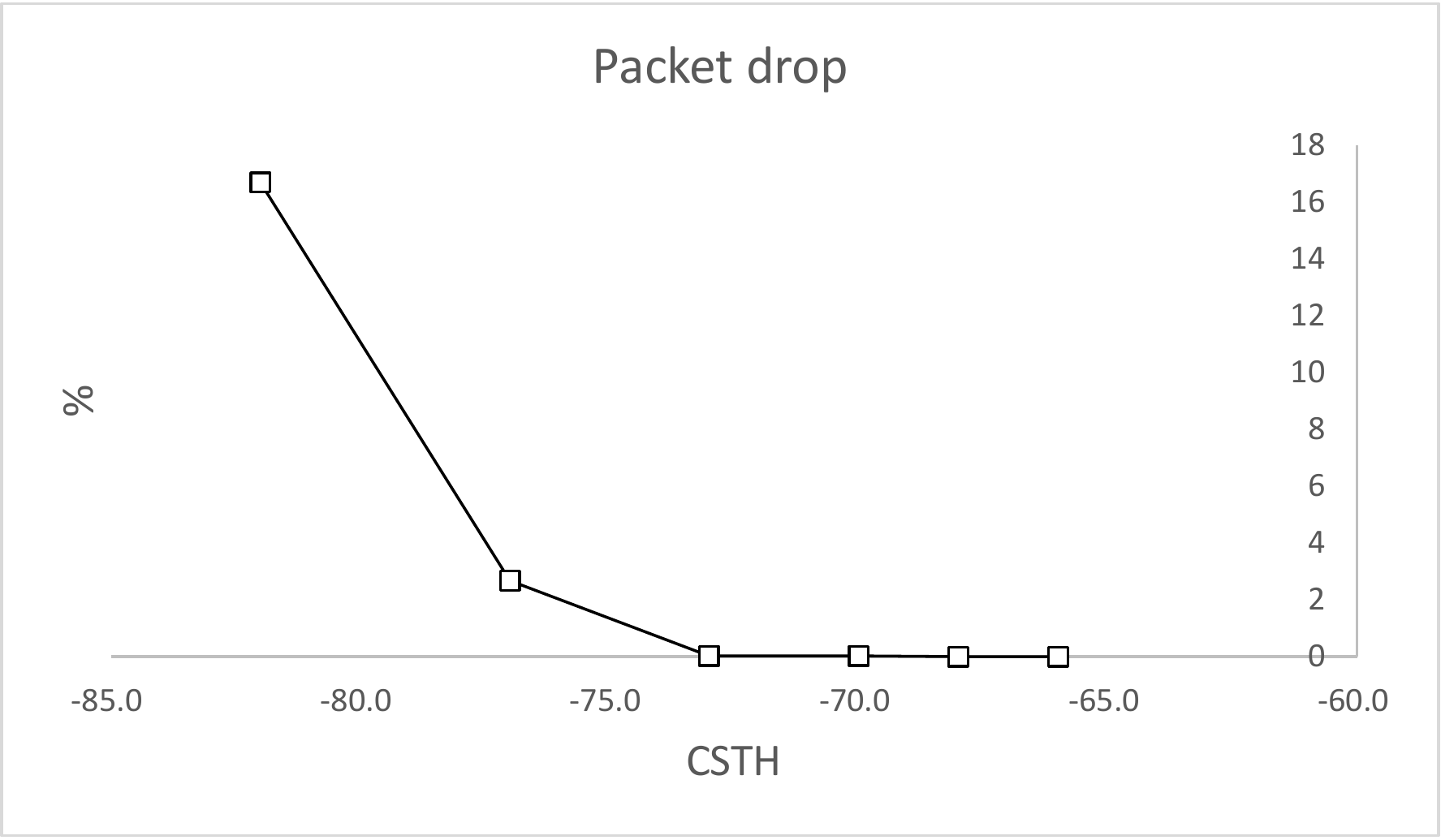}
        \caption{Packet drop with CSTH increased to 4800
	packets/sec.\label{packetdropcsth}}
    \end{minipage}
\end{figure*}

We note that, in our simulations, we assume that after a collision is detected,
colliding packets are retransmitted if the retry limit is not reached.
Otherwise, the packets are dropped.  Fig.~\ref{packetdroph} shows that,
in the presence of hidden terminals, the packet drop is very low at low range of offered load values (2400 packets/sec).  However, with the increase of packet
arrival rate, packet drop increases and becomes as high as 16\% at an
arrival rate of 4800 packets/sec in the presence of hidden terminals.
Thus, hidden terminals severely impact the performance of the network
at high load condition.  At CSTH of -73dBm the packet drop is almost
zero as shown from Fig.~\ref{packetdropcsth}.

\begin{figure*}[t!]
    \centering
    \begin{minipage}[b]{0.33\textwidth}
        \centering
        \includegraphics[width=0.95\textwidth]{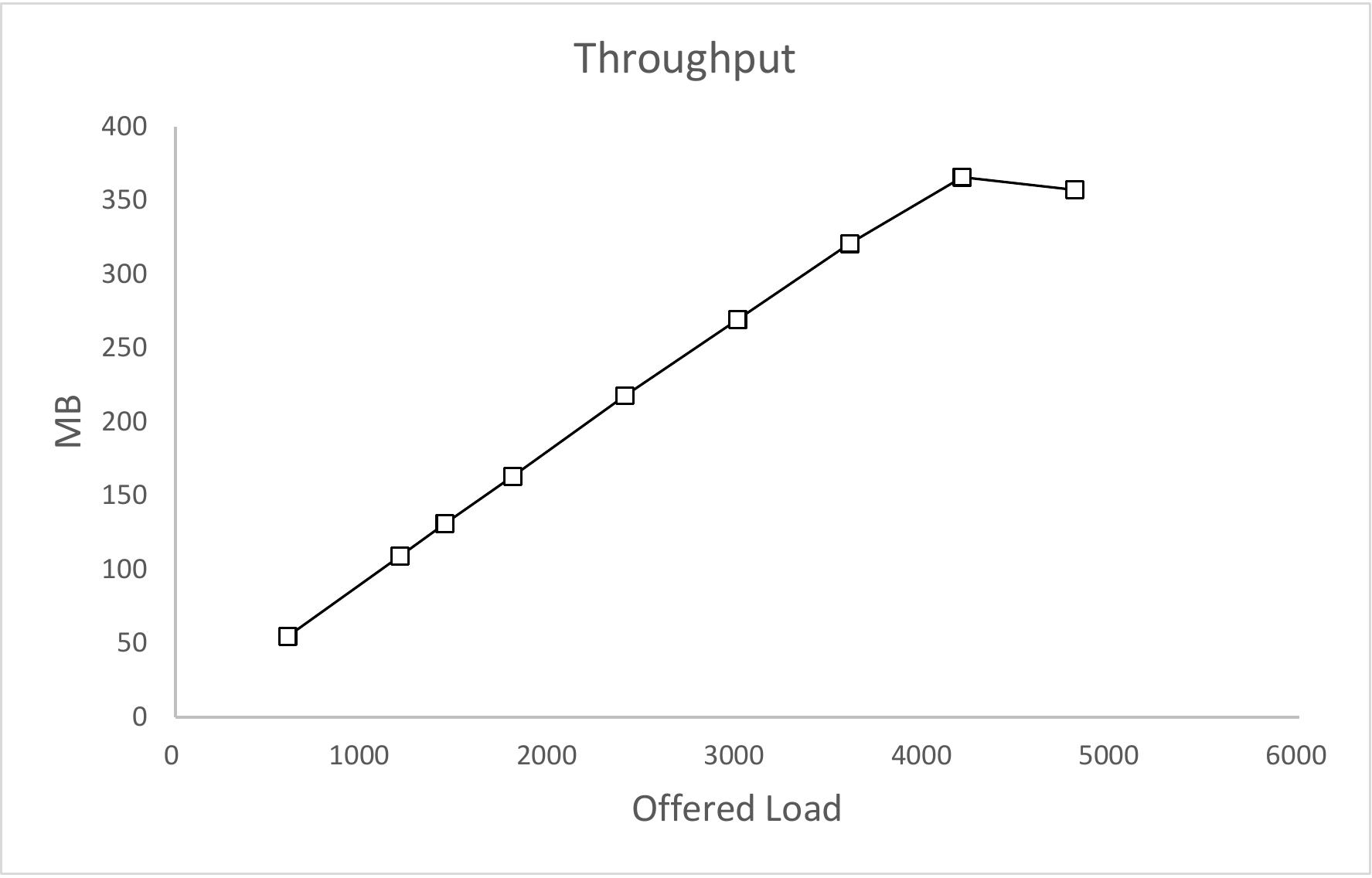}
        \caption{Network throughput, hidden node present.\label{throughputh}}
    \end{minipage}%
    ~ 
    \begin{minipage}[b]{0.33\textwidth}
        \centering
        \includegraphics[width=0.95\textwidth]{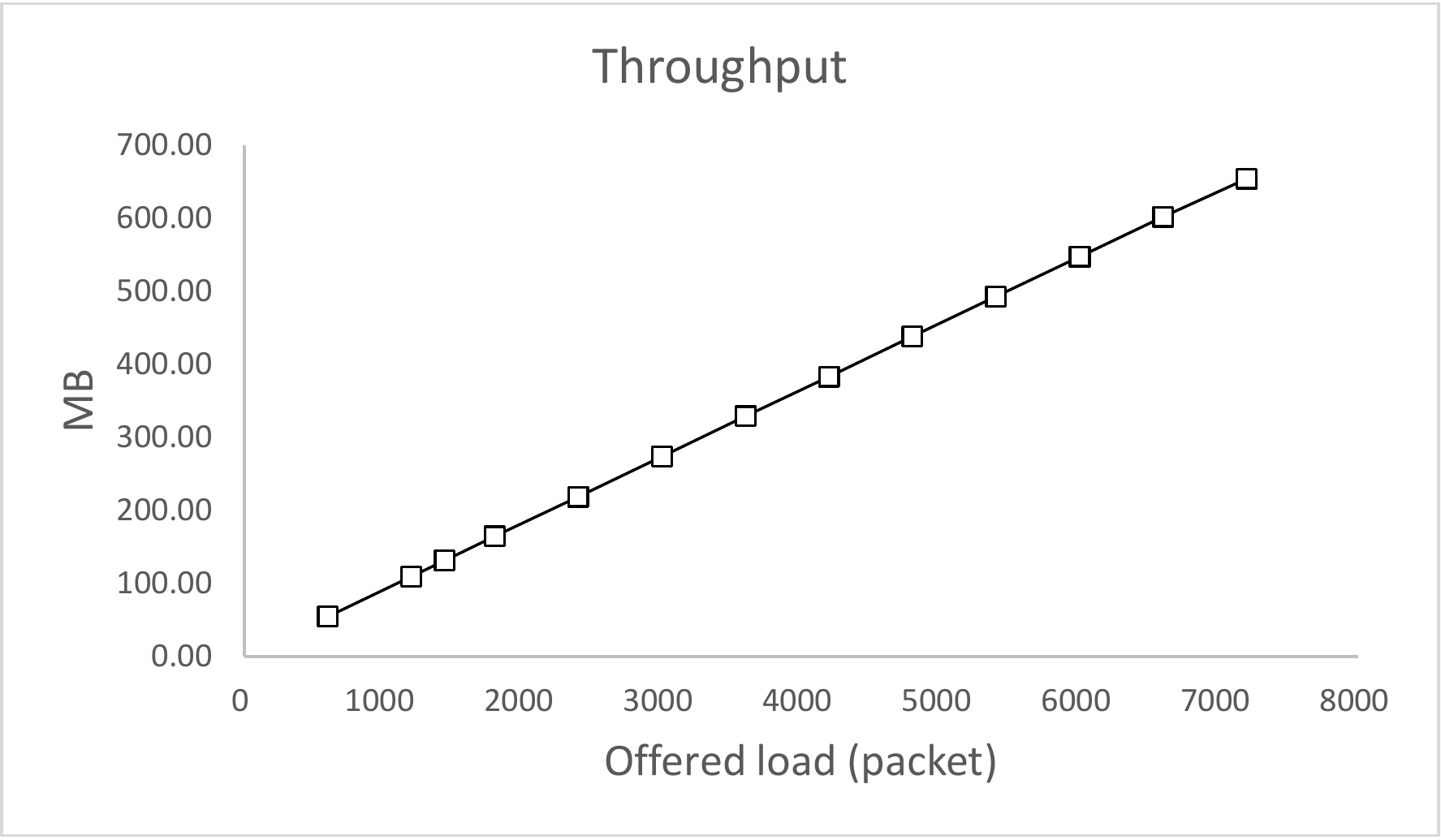}
        \caption{Network throughput, no hidden node present.\label{throughputn}}
    \end{minipage}%
    ~ 
    \begin{minipage}[b]{0.33\textwidth}
        \centering
        \includegraphics[width=0.95\textwidth]{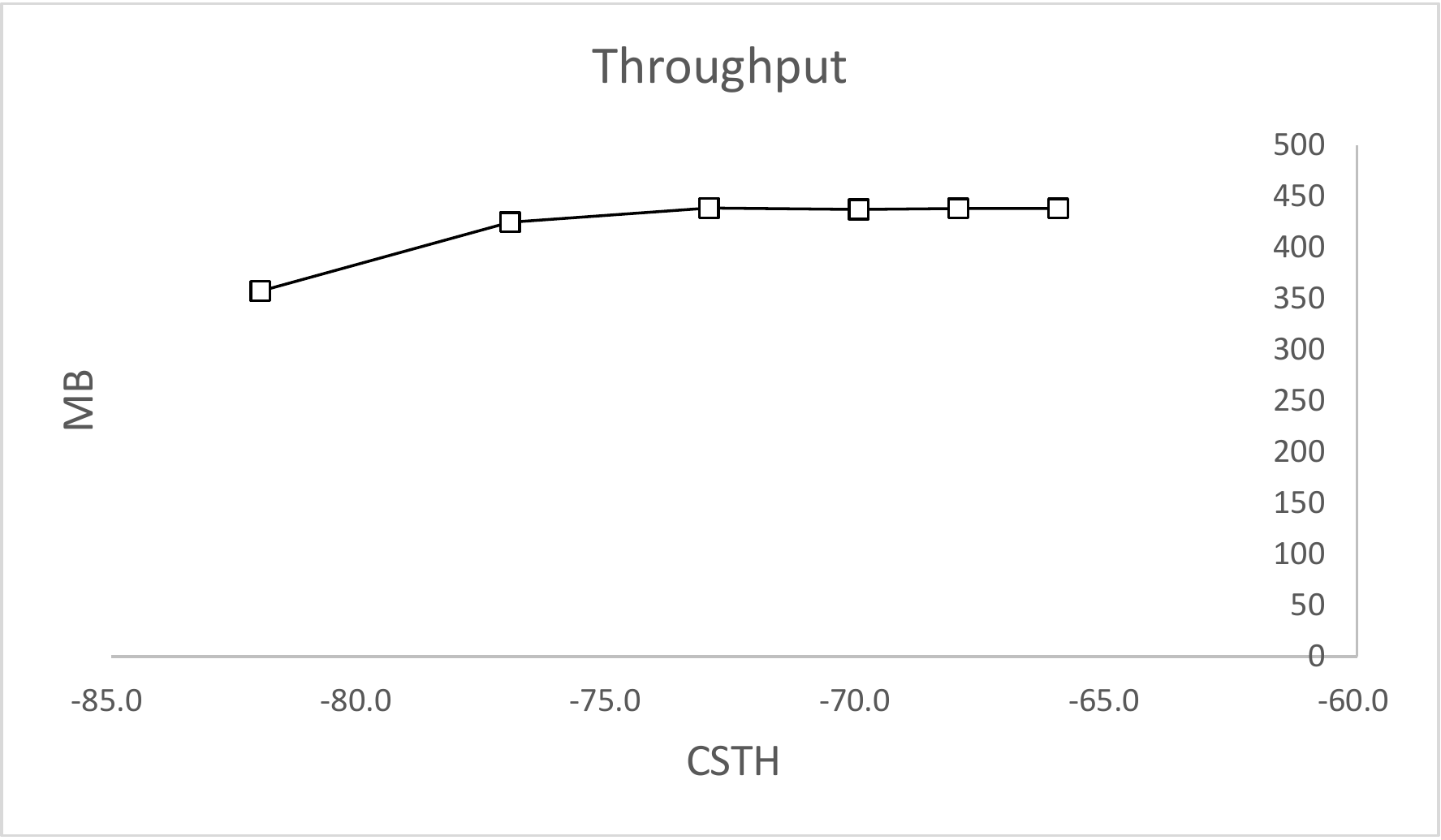}
        \caption{Network throughput with CSTH increased to 4800
	packets/sec.\label{throughputcsth}}
    \end{minipage}
\end{figure*}

Network throughput is severely impacted by collisions due to hidden
terminals.  From Fig.~\ref{throughputh} we observe that, at an arrival
rate of 4200 packets/sec, we achieve highest network throughput of about q366
Mbps. The network throughput at corresponding arrival rate without
hidden terminal is 383 Mbps. This deviation of throughput due to
hidden terminal is due to longer waiting time, longer backoff time and
larger number of retransmissions. 

If we further increase the packet arrival rate (4800 packets/sec), the
network throughput drops to 357 Mbps as all packets can not be
transmitted due to collision and eventually, packets are dropped.
However, without hidden terminal problem, network throughput linearly
increases with the increase of packet arrival rate as shown in
Fig.~\ref{throughputn}.  The network throughput at an arrival rate of
4800 packets/sec reaches about 438 Mbps without hidden nodes which is
23\% higher than the throughput achieved with hidden terminal
scenario.  

We also observe that in absence of
hidden terminals, the network can sustain larger packet arrival rate
of even 7200 packets/sec and a high throughput of 655 Mbps can be
achieved. With the increase of CSTH from -82dBm to -73dBm at an
arrival rate of 4800 packets/sec, the throughput of the network with
hidden terminals increases from 357 Mbps to 438 Mbps which is about the same as the
throughput obtained in the network without hidden terminals.

\begin{figure*}[t!]
    \centering
    \begin{minipage}[b]{0.475\textwidth}
        \centering
        \includegraphics[width=0.95\textwidth]{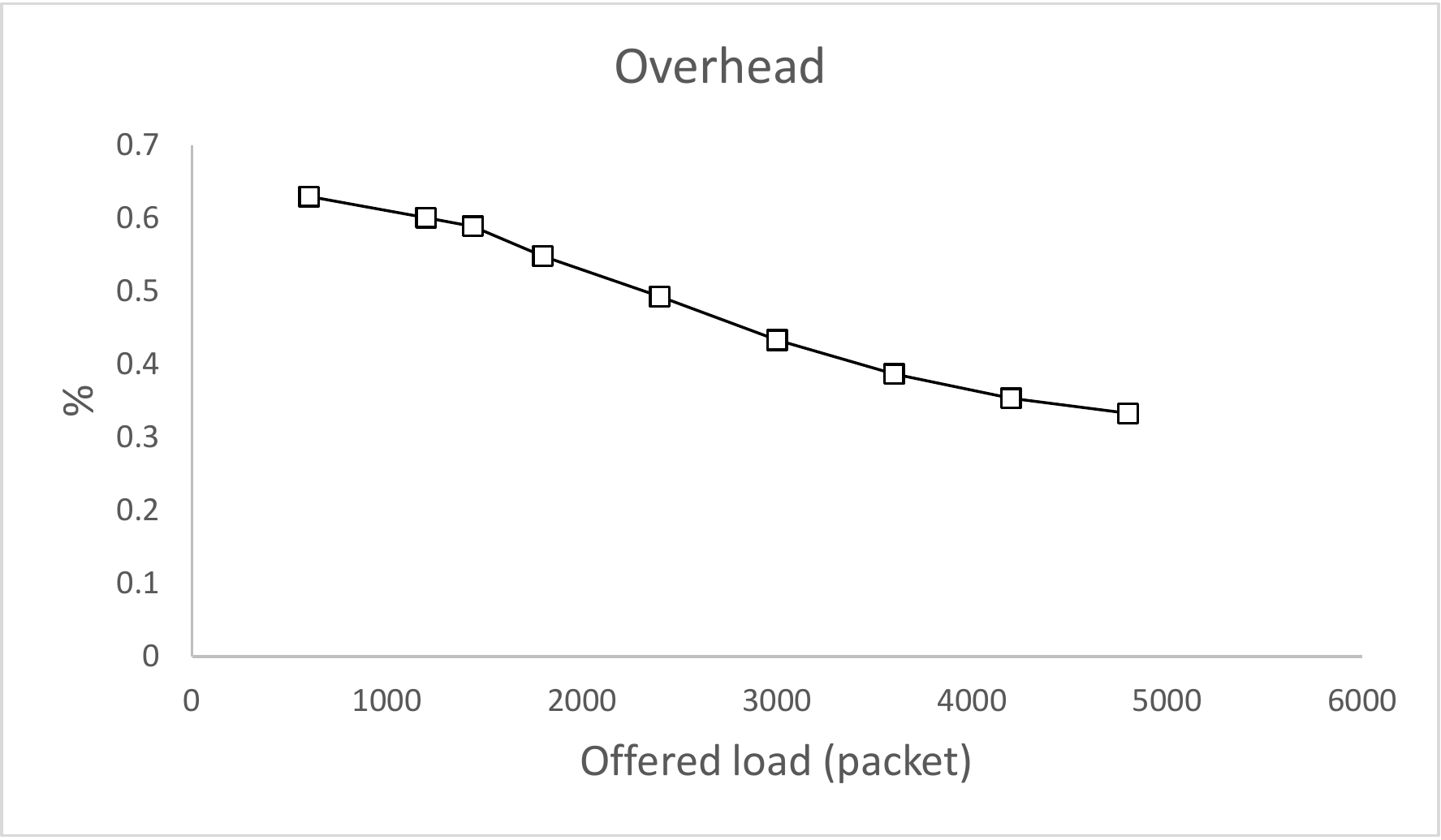}
        \caption{MAC overhead, hidden node present.\label{overheadh}}
    \end{minipage}%
    ~ 
    \begin{minipage}[b]{0.475\textwidth}
        \centering
        \includegraphics[width=0.95\textwidth]{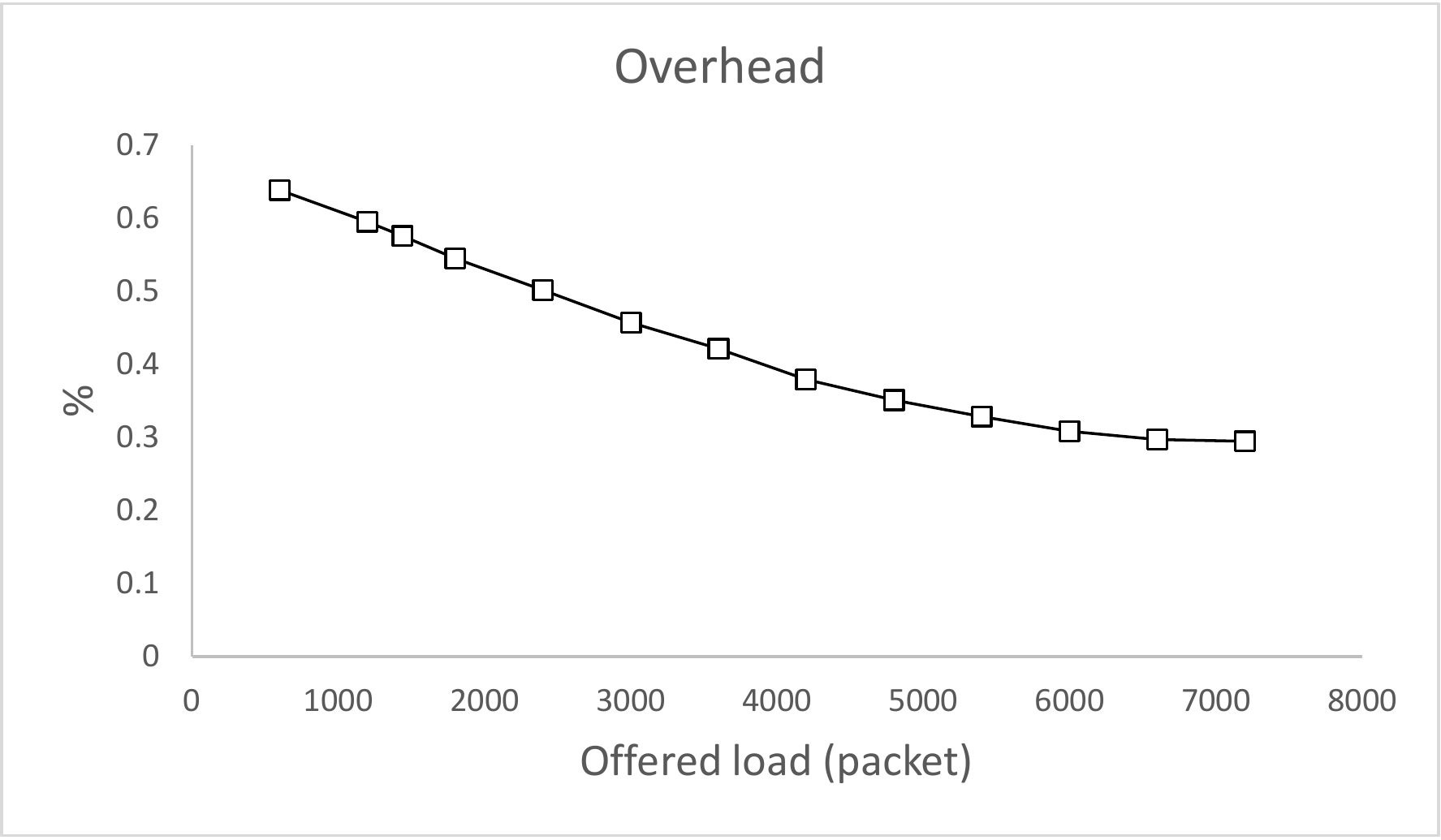}
        \caption{MAC overhead, no hidden node present\label{overheadn}.}
    \end{minipage}
\end{figure*}

Overhead is calculated as the ratio of time required to transmit
control signals to data packet. Figs.~\ref{overheadh} and \ref{overheadn} show a comparison
of MAC overhead in presence of hidden terminal and without hidden
terminal scenarios. At low packet arrival rate, most of the time STAs
initiate RTS/CTS controlled single user transmission individually. As
a result, the overhead for a data packet is much higher. As the packet
arrival rate increases, more STAs start MU transmission where data
packets and control signals are simultaneously transmitted from
multiple STAs. As a result overhead cost gradually decreases with the
increase in packet arrival rate. Due to increase of retransmission of
control signals in hidden terminal scenario, MAC overhead is higher
for the network with hidden terminals.

\begin{figure*}[t!]
    \centering
    \begin{minipage}[b]{0.475\textwidth}
        \centering
        \includegraphics[width=0.95\textwidth]{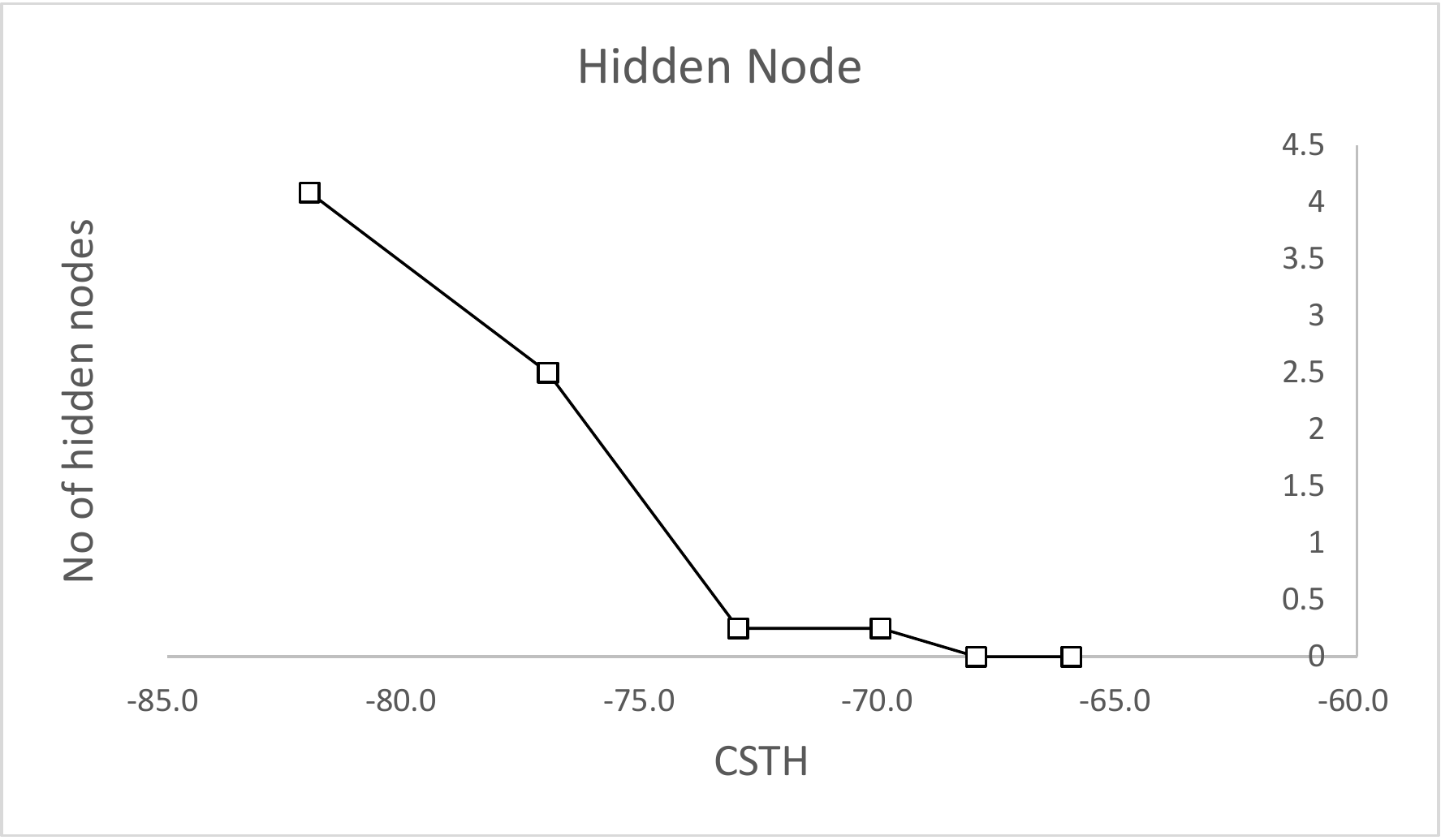}
        \caption{Average number of hidden nodes per STA.\label{hiddennodecsth}}
    \end{minipage}%
    ~ 
    \begin{minipage}[b]{0.475\textwidth}
        \centering
        \includegraphics[width=0.95\textwidth]{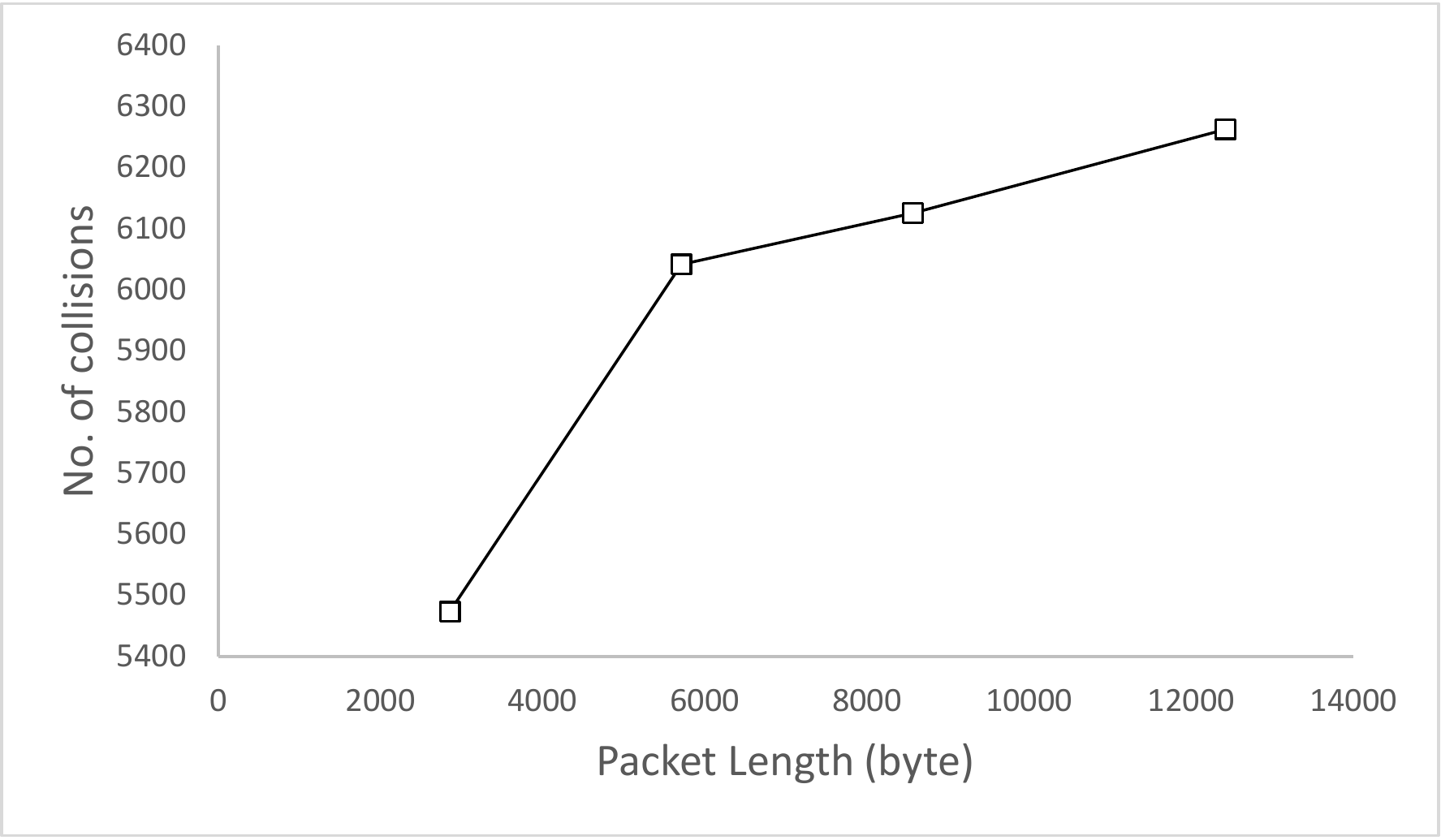}
        \caption{Variation of collision probability.\label{collpacketsize}}
    \end{minipage}
\end{figure*}

In the existing CSTH (-82dBm), each STA on an average sees four hidden
terminals during transmission. With the increase of CSTH from -82 dBm
to -73dBm, the average number of hidden terminals seen by a node
reduces to 0.25 as shown in Fig.~\ref{hiddennodecsth}.
Fig.~\ref{collpacketsize} shows the number of collisions
as a function of packet size. As the packet size decreases, the
transmission time required by each packet decreases. As a result,
nodes get frequent transmission opportunities and the collision probability decreases. However, this reduction
is not
significant because data packets do not suffer collision because, as soon as
AP transmits CTS, the NAV of all other STAs are updated which
eliminates hidden node problem for data packets.

\section{Conclusion}\label{conclusion}

We discussed the impact of hidden node problem in uplink transmission
in a heterogeneous network. The simulation result showed significant
throughput degradation due to the presence of hidden nodes in
coexisting network. We strongly advocate in favour of increasing
carrier sensing threshold of STAs by 10 dB during association with HE
AP which not only increase network capacity but also reduce collision
probability due to hidden node. We proposed some modification to the
draft recommendation to reduce the inter BSS interference arising from
asymmetrical transmission radius of AP and STAs.

\section{Acknowledgments}\label{sec11}

Research presented here was in part supported through Canada's National Science and Engineering Research Council (NSERC) Discovery Grants.

\bibliographystyle{unsrt}

\end{document}